\documentclass[12pt,a4paper,dvips]{article}

\usepackage{a4p}
\usepackage{cite}
\usepackage{graphicx}
\usepackage{physics}
\usepackage{l3_title}
\usepackage{rotating}
\usepackage{amssymb}

\date{October 24, 2003}
\preprint{2003-068}
\journalname{Physics Letters B}

\newlength{\capindent}
\newlength{\capwidth}
\newlength{\figwidth}
\newcommand{\icaption}[2][!*!,!]{\hspace*{\capindent}%
  \begin{minipage}{\capwidth}
    \ifthenelse{\equal{#1}{!*!,!}}%
      {\caption{#2}}%
      {\caption[#1]{#2}}
  \end{minipage}}

\begin{document}

\begin{titlepage}

  \title{Single- and Multi-Photon Events with Missing Energy \\
    in {\boldmath \epem} Collisions at LEP}

  \author{The L3 Collaboration}

%
% The abstract
%
  \begin{abstract}

Single- and multi-photon events with missing energy are selected in
619 pb$^{-1}$ of data collected by the L3 detector at LEP at
centre-of-mass energies between 189~\GeV\ and 209~\GeV.  The cross
sections of the process $\mathrm{e}^+\mathrm{e}^- \rightarrow \nu
\bar{\nu} \gamma (\gamma) $ are found to be in agreement with the
Standard Model expectations, and the number of light neutrino species
is determined, including lower energy data, to be $N_\nu = 2.98 \pm
0.05 \pm 0.04$.  Selection results are given in the form of tables
which can be used to test future models involving single- and
multi-photon signatures at LEP.  These final states are also predicted
by models with large extra dimensions and by several supersymmetric
models.  No evidence for such models is found. Among others, lower
limits between 1.5~\TeV\ and 0.65~\TeV\ are set, at 95\% confidence
level, on the new scale of gravity for the number of extra dimensions
between 2 and 8.

 \end{abstract} 

\submitted

\end{titlepage}

\def \nngg {$\nu_{\mathrm{e}}\bar\nu_{\mathrm{e}}\gamma$}
\def\gravin{\ensuremath{\tilde\mathrm{G}}}%
\def\DELTAM{\ensuremath{\Delta m}}%
\def\Msel{\ensuremath{m_{\tilde\e_\mathrm{L}}}}%
\def\Mser{\ensuremath{m_{\tilde\e_\mathrm{R}}}}%
\def\Mselr{\ensuremath{m_{\tilde\e_\mathrm{L,R}}}}%
\def\Mchi{\ensuremath{m_{\tilde\chi^0_1}}}%
\def\Mchii{\ensuremath{m_{\tilde\chi^0_2}}}%
\def\Mcha{\ensuremath{m_{\tilde\chi^\pm_1}}}%
\def\Msnu{\ensuremath{m_{\tilde\nu}}}%
\def\Msml{\ensuremath{m_{\tilde\mu_\mathrm{L}}}}%
\def\Mstat{\ensuremath{m_{\tilde\tau_2}}}%
\def\MP{\ensuremath{m_{\mathrm{P}}}}%
\def\MG{\ensuremath{m_{\gravin}}}%
\def\susyl#1{\ensuremath{\tilde{#1}_\mathrm{L}}}%
\def\susyr#1{\ensuremath{\tilde{#1}_\mathrm{R}}}%
\def\susylr#1{\ensuremath{\tilde{#1}_\mathrm{L,R}}}%
\def\tanb{\ensuremath{\tan \beta}}%

%
%%%%%%%%%%%%%%%%%%%%%%%%%%%%%%%%%%%%%%%%%%%%%%%%%%%%%%%%%%%%%%%%%%%%%%%%%%%%%%%
\section{Introduction}
%%%%%%%%%%%%%%%%%%%%%%%%%%%%%%%%%%%%%%%%%%%%%%%%%%%%%%%%%%%%%%%%%%%%%%%%%%%%%%%
%

In the Standard Model of the electroweak
interactions~\cite{standard_model} single- or multi-photon events with
missing energy are produced via the reaction \epem\ \ra\
\nnbar\gam(\gam) which proceeds through $s$-channel Z exchange and
$t$-channel W exchange.  The majority of such events are due to initial
state radiation (ISR) from the incoming electrons and
positrons\footnote{A small fraction of photons originates from the
$t$-channel W boson fusion in the \ee\ \ra\ \nngg(\gam) process.}.  The
distribution of the recoil mass to the photon system, $M_{rec}$, is
expected to peak around the Z mass in the $s$-channel, whereas ISR
photons from the $t$-channel W exchange are expected to have a relatively 
flat energy distribution, peaked at low energies~\cite{bardin}.

This Letter describes L3 results from the highest energy and
luminosity LEP runs and improves upon and supersedes previous
publications~\cite{papgg99}.  Other LEP experiments also reported
similar studies~\cite{paplepgg}.  The cross section measurement of the
\epem\ \ra\ \nnbar\gam(\gam) process is presented, as well as the
direct measurement of the number of light neutrino species. 
 Selection
results are also given in the form of tables which can be used to test
future models involving single- and multi-photon signatures at LEP.

The selected events are used to search for manifestations of Physics
beyond the Standard Model, such as extra dimensions and Supersymmetry
(SUSY). Models with large extra dimensions~\cite{qgrav1} predict a
gravity scale, $M_D$, as low as the electroweak scale, naturally
solving the hierarchy problem. Gravitons, G, are then produced in
\epem\ collisions through the process \epem\ \ra\ \gam G, and escape
detection, leading to a single-photon signature. Different mechanisms
are suggested for symmetry breaking in SUSY models~\cite{susy}, which
imply three different scenarios: ``superlight'', ``light'' and ``heavy''
gravitinos, \gravin, with several single- or multi-photon and missing
energy signatures. Results of generic searches for \ee \ra\ XY \ra\
YY\gam\ and \ee \ra\ XX \ra\ YY\gam\gam, where X and Y are new 
neutral invisible particles, are also discussed.

The main variables used in the selection of single- and multi-photon
events are the photon energy, $E_\gamma$, polar angle,
$\theta_{\gamma}$, and transverse momentum, $P_t^\gamma$.  Three event
topologies are considered:

\begin{itemize}

\item High energy single-photon: a photon with $14^{\circ} <
 \theta_{\gamma} < 166^{\circ}$ and $P_t^\gamma > 0.02\sqrt{s}$. There
 should be no other photon with $E_\gamma > 1 \GeV$.

\item Multi-photon: at least two photons with $E_\gamma > 1 \GeV$,
with the most energetic in the region $14^{\circ} < \theta_{\gamma} <
166^{\circ}$ and the other in the region $11^{\circ} < \theta_{\gamma}
< 169^{\circ}$.  The transverse momentum of the multi-photon system
should satisfy $P^{\gamma\gamma}_t > 0.02\sqrt{s}$.

\item Low energy single-photon: a photon with $43^{\circ} <
\theta_{\gamma} < 137^{\circ}$ and $ 0.008\sqrt{s} < P_t^\gamma <
0.02\sqrt{s}$.  There should be no other photon with $E_\gamma >
1 \GeV$.

\end{itemize}

The inclusion of the low energy single-photon sample significantly
increases the sensitivity of the searches for extra dimensions and
pair-produced gravitinos.

%
%%%%%%%%%%%%%%%%%%%%%%%%%%%%%%%%%%%%%%%%%%%%%%%%%%%%%%%%%%%%%%%%%%%%%%%%%%%%%%%
\section{Data and Monte Carlo Samples}
%%%%%%%%%%%%%%%%%%%%%%%%%%%%%%%%%%%%%%%%%%%%%%%%%%%%%%%%%%%%%%%%%%%%%%%%%%%%%%%
%

Data collected by the L3 detector \cite{L3-DETECTOR} at LEP in the
years from 1998 through 2000 are considered. They correspond to an
integrated luminosity of 619~\pb\ at centre-of-mass energies
$\sqrt{s}= 188.6 -209.2 \GeV$, as detailed in Table~\ref{tab:lumeff}.

The following Monte Carlo generators are used to simulate Standard
Model processes: {\tt KKMC}~\cite{KKMC} for \epem\ \ra\
\nnbar\gam(\gam), {\tt GGG}~\cite{GGG} for \epem\ \ra\ \gam\gam(\gam),
{\tt BHWIDE}~\cite{BHWIDE} and {\tt TEEGG}~\cite{teegg} for large and
small angle Bhabha scattering, respectively, {\tt DIAG36}~\cite{DIAG}
for \epem\ \ra\ \epem\epem\ and {\tt EXCALIBUR}~\cite{excalibur_new}
for \epem\ \ra\ \epem\nnbar. The predictions of {\tt KKMC} are checked
with the {\tt NUNUGPV}~\cite{NUNUGPV} generator.  SUSY processes are
simulated with the {\tt SUSYGEN}~\cite{SUSYGEN2.2} Monte Carlo
program, for SUSY particles with masses up to the kinematic limit.

The L3 detector response is simulated using the {\tt GEANT}
program~\cite{geant}, which describes effects of energy loss, multiple
scattering and showering in the detector.  Time-dependent detector
inefficiencies, as monitored during the data taking period, are
included in the simulation.

%
%%%%%%%%%%%%%%%%%%%%%%%%%%%%%%%%%%%%%%%%%%%%%%%%%%%%%%%%%%%%%%%%%%%%%%%%%%%%%%
\section{Event Selection}
%%%%%%%%%%%%%%%%%%%%%%%%%%%%%%%%%%%%%%%%%%%%%%%%%%%%%%%%%%%%%%%%%%%%%%%%%%%%%%
%

Electrons and photons are reconstructed in the BGO crystal
electromagnetic calorimeter (ECAL).  It is accurately calibrated using
an RFQ accelerator~\cite{rfq} and has an energy resolution
$\sigma(E)/E = 0.035/\sqrt{E} \oplus 0.008$.  Its barrel region
subtends the polar angle range $43^\circ < \theta < 137^\circ$ while
the endcap regions subtend the ranges $10^\circ < \theta < 37^\circ$ and
$143^\circ < \theta < 170^\circ$. The region between the barrel and
the endcaps is instrumented with a lead and scintillator fiber
electromagnetic calorimeter (SPACAL), which is used as a veto
counter to ensure the hermeticity of the detector.  The fiducial
volume of the tracking chamber (TEC), used to discriminate between photons
and electrons, is $14^\circ < \theta < 166^\circ$.

Photon candidates are required to have an energy greater than 1~\gev\
and the shape of their energy deposition must be consistent with an
electromagnetic shower.  Bhabha  and \epem\ \ra\ \gam\gam(\gam)
events that are fully contained in the ECAL are used to check the
particle identification efficiency and the energy resolution.

Single- and multi-photon events are accepted by calorimetric triggers
monitored with a control sample of single-electron events. These are
radiative Bhabha scattering events where one electron and a photon
have a very low polar angle, and only a low energy electron is
scattered at a large polar angle. They are accepted by a dedicated
independent trigger requiring the coincidence of a charged track and a
cluster in one of the luminosity monitors. Figure~\ref{fig:sel_plots}a
shows the trigger efficiency as a function of the ECAL shower
energy. In the barrel, it rises sharply at the energy threshold of a first trigger
and reaches a plateau mainly
determined by the presence of inactive
channels~\cite{l3enertrigg_l3firstsgpaper}. With increasing energy
additional triggers become active, resulting in a second 
 threshold rise and a final plateau at efficiencies
of $92.3\pm0.6\%$ in the barrel and $95.4\pm0.4\%$ in the endcaps.
As the cross section of single-electron
production decreases rapidly with the single-electron energy, the
trigger performance study at high energies is complemented by studying
Bhabha events selected using calibration data at the Z
peak.

%
%%%%%%%%%%%%%%%%%%%%%%%%%%%%%%%%%%%%%%%%%%%%%%%%%%%%%%%%%%%%%%%%%%%%%%%%%%%%%%
\subsection{High Energy Single-Photon Selection}
%%%%%%%%%%%%%%%%%%%%%%%%%%%%%%%%%%%%%%%%%%%%%%%%%%%%%%%%%%%%%%%%%%%%%%%%%%%%%%
%

The selection of high energy single-photon events requires only one
photon candidate in the barrel or endcaps with transverse momentum
$P_t^\gamma~>~0.02\sqrt{s}$.  The energy not assigned to the
identified photons must be less than 10~\gev{} and the energy measured
in the SPACAL must be less than 7~\gev.  There must be no tracks in
the muon chambers and at most one ECAL cluster not identified as a
photon is allowed in the event.  Electron candidates are removed by
requiring that no charged track reconstructed in the TEC matches the
ECAL cluster.

The probability of photon conversion in the beam pipe and in the
silicon microvertex detector is about 5\% in the barrel region and
increases rapidly at low polar angles, reaching about 20\% at $\theta
\approx 20^\circ$.  To improve the selection 
efficiency in the presence of converted
photons, the cut on the TEC tracks is released for events with $M_{rec} =
80-110 \GeV$ in the barrel and $M_{rec} =80-140 \GeV$ in the
endcaps. Photon candidates in the barrel region with $M_{rec}$ outside
this range are also accepted if they have two matching tracks with an
azimuthal opening angle $\Delta\Phi_{tracks}<15^\circ$.  The
distribution of $\Delta\Phi_{tracks}$ for photons accepted by this cut
is presented in Figure~\ref{fig:sel_plots}b.

To reduce background from radiative Bhabha events at low polar angles and
from the process \epem\ \ra\ \gam\gam(\gam), events with a transverse
momentum less than 15~\gev{} are rejected if an energy cluster is
observed in the forward calorimeters covering an angular range of
$1.5^\circ-10^\circ$, with an acoplanarity\footnote{ Defined as the
complement of the angle between the projections in the plane
perpendicular to the beam axis.}  with the most energetic photon less
than 30$^\circ$.  Furthermore, if a photon is detected with an
acoplanarity less than 15$^\circ$ with a hadron calorimeter cluster,
the energy of this cluster must be less than 3~\gev{}.

To reject cosmic ray background, no muon track segments are allowed in
the event for photons with energy less than 40~\GeV. If photons are
more energetic, their ECAL showers leak into the time-of-flight system
and its signals are required to be in time with the beam crossing
within $\pm$5~ns.  Furthermore, an event is rejected if more than 20
hits are found in the central tracking chamber in a 1~cm road between
any pair of energy depositions in the ECAL.  The cosmic ray background
in the event sample is estimated from studies of out-of-time events
and amounts to 0.2\%.

The noise in various subdetectors is studied using events randomly
triggered at the beam crossing time. The resulting efficiency loss is
0.8\%, and the Monte Carlo predictions are scaled accordingly.

\begin{figure}
  \begin{center}
\begin{tabular}{ll}
    \includegraphics[width=0.45\textwidth ]{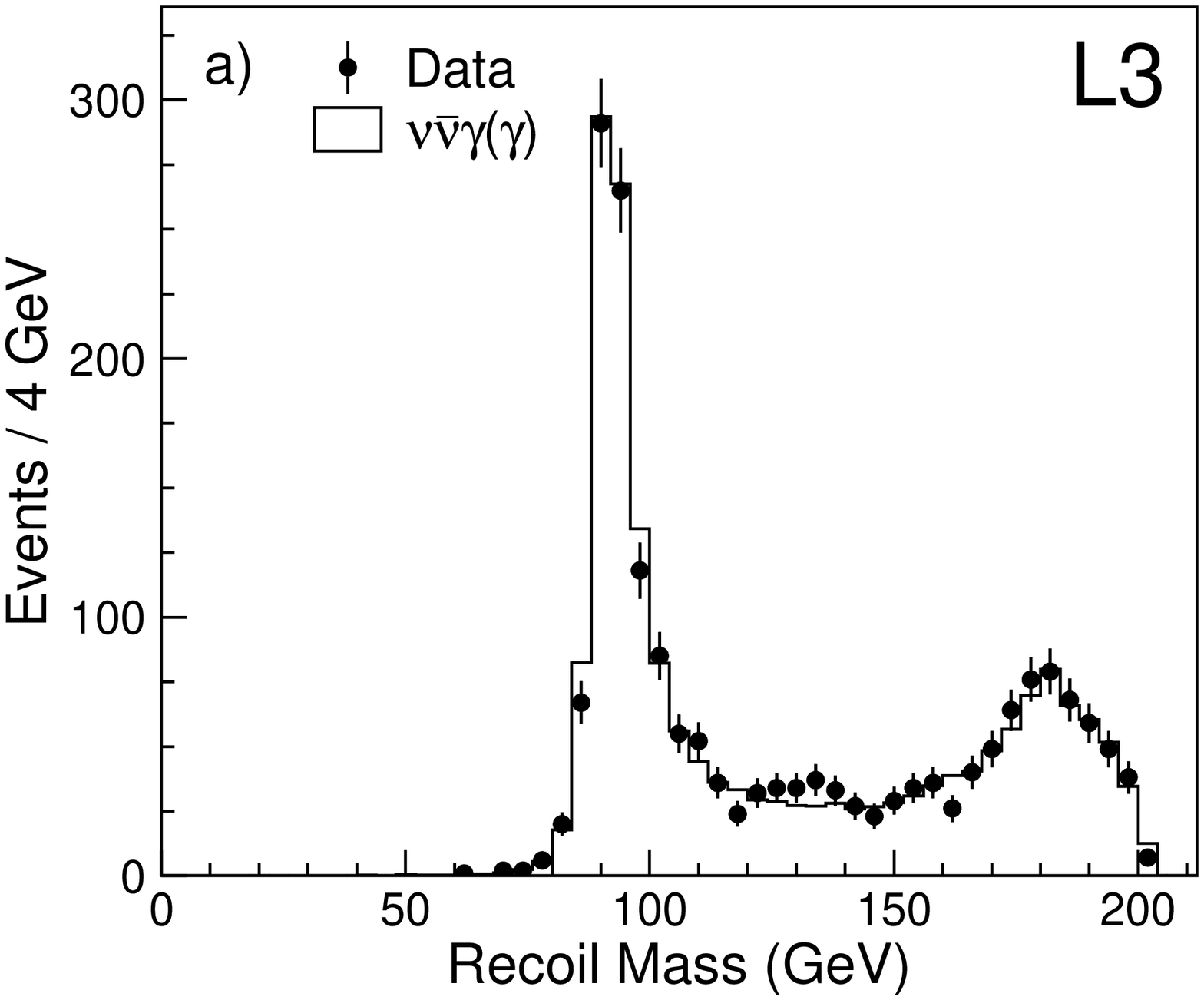}& 
    \includegraphics[width=0.45\textwidth ]{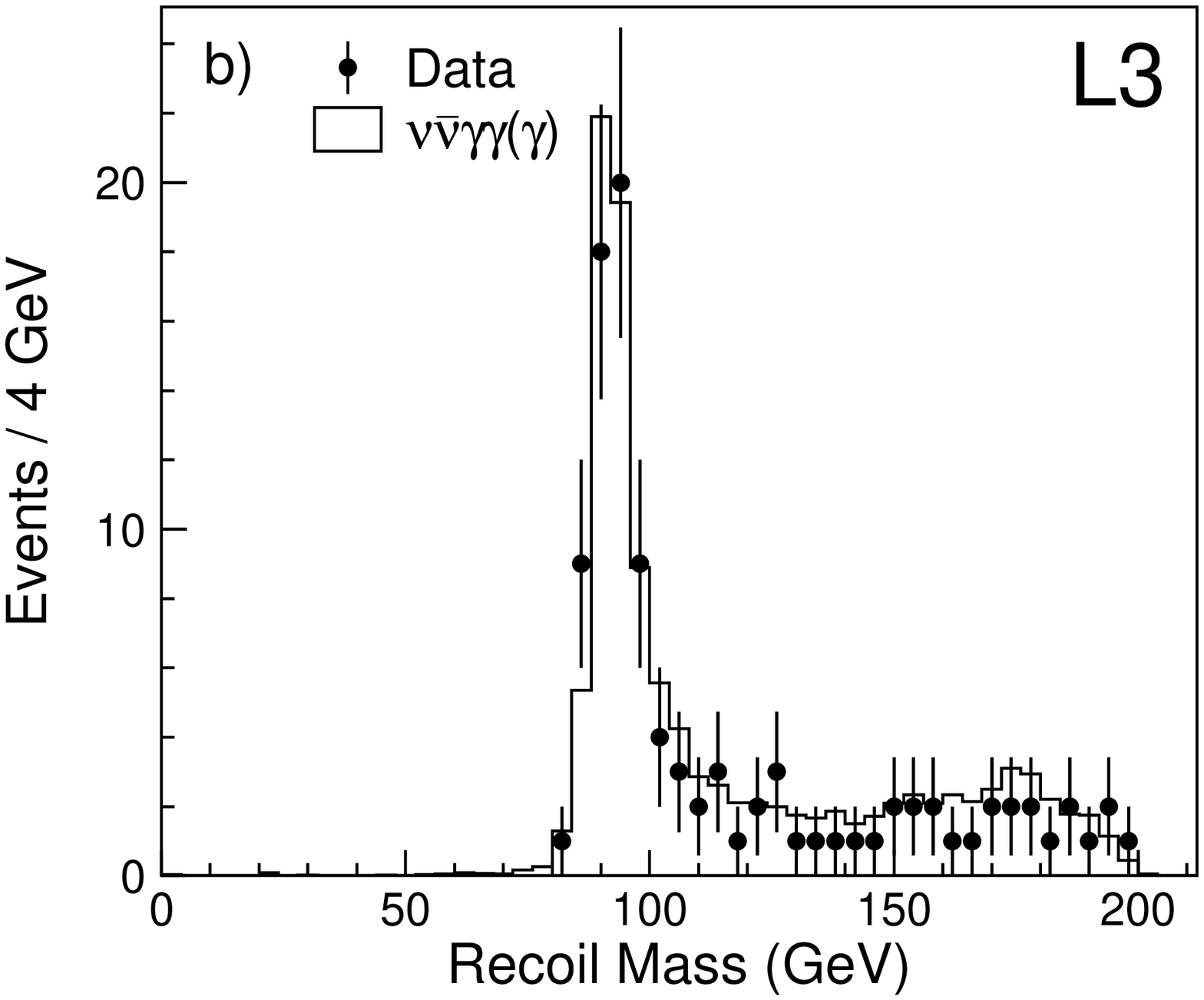}\\
    \includegraphics[width=0.45\textwidth ]{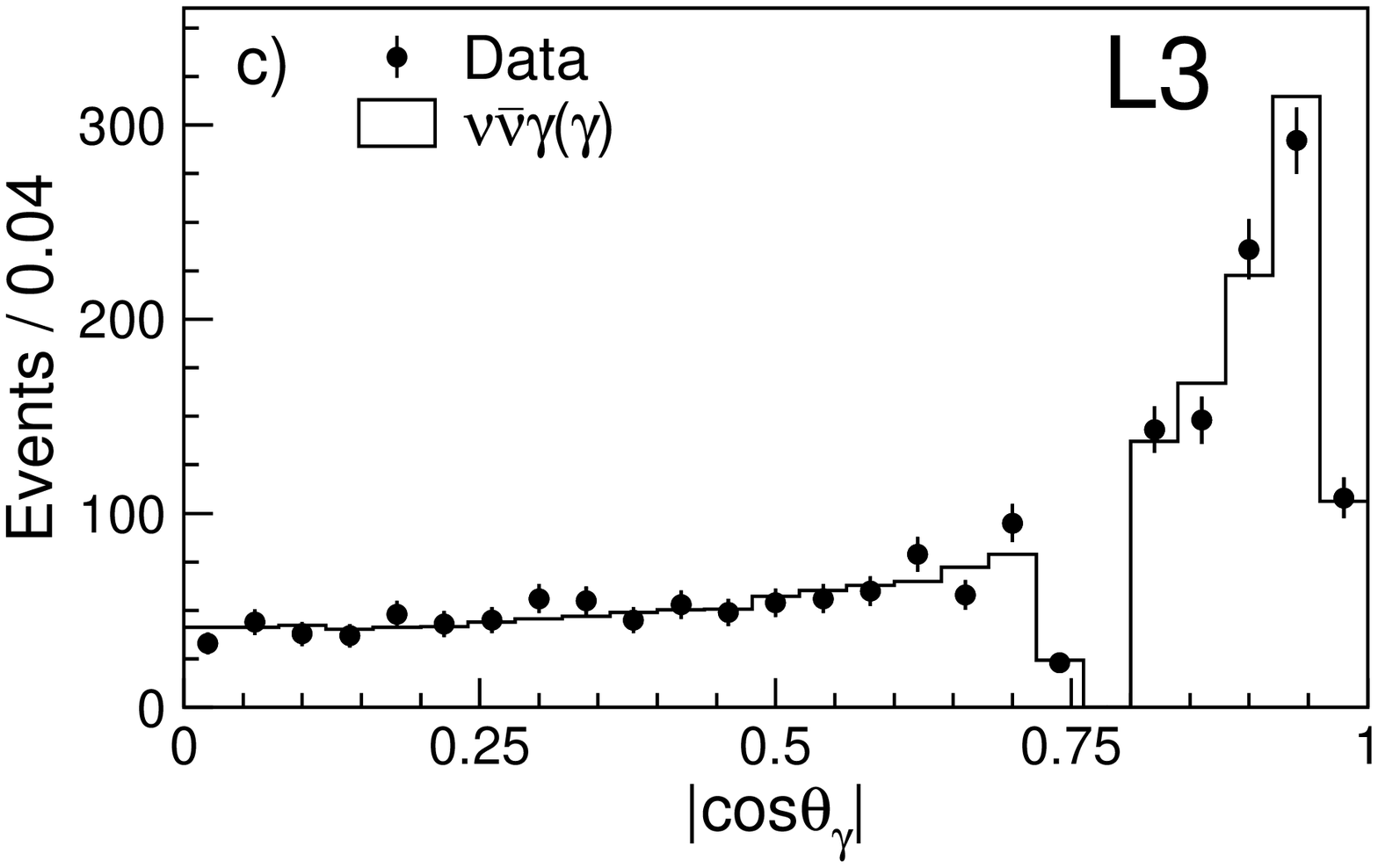}& 
    \includegraphics[width=0.45\textwidth ]{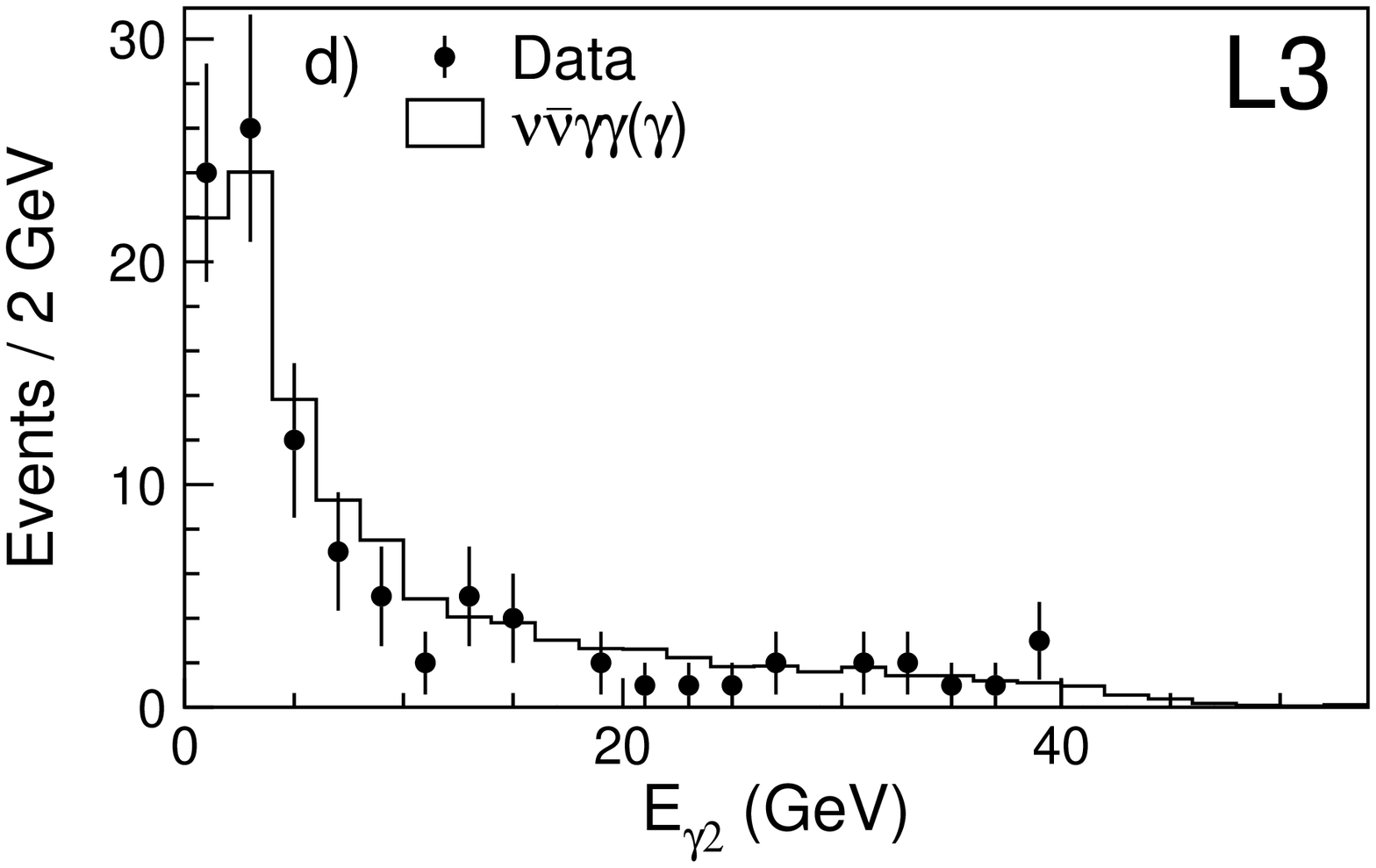}\\
 \end{tabular}
    \icaption{\label{fig:egamma} Distributions of a) the recoil mass
and c) the polar angle for the high energy single-photon events and of
b) the recoil mass and d) the energy of the second most energetic
photon for the multi-photon sample.  }
  \end{center}
\end{figure}

In total, 1898 events are selected in data with 1905.1 expected from
Monte Carlo.  The purity of the selected 
$\epem\rightarrow\nnbar\gam(\gam)$ sample
is estimated to be 99.1\%, with the main background coming from
radiative Bhabha events and from the
$\epem\rightarrow\gamma\gamma(\gamma)$ process.
Figures~\ref{fig:egamma}a and~\ref{fig:egamma}c show the distributions
of $M_{rec}$ and $|\cos{\theta_\gamma}|$. The numbers of events
selected at different values of $\sqrt{s}$ are listed in
Table~\ref{tab:sgamsel}, together with the Standard Model
expectations. The efficiencies of the selection and the numbers of
observed and expected events are given in Table~\ref{tab:highener_sg}
in bins of $M_{rec}$ and $|\cos\theta_\gamma|$.

%
%%%%%%%%%%%%%%%%%%%%%%%%%%%%%%%%%%%%%%%%%%%%%%%%%%%%%%%%%%%%%%%%%%%%%%%%%%%%%%
\subsection{Multi-Photon Selection}
%%%%%%%%%%%%%%%%%%%%%%%%%%%%%%%%%%%%%%%%%%%%%%%%%%%%%%%%%%%%%%%%%%%%%%%%%%%%%%
%

Multi-photon candidates should have at least two photons with
energy above 1~\GeV\ and a global transverse momentum $P^{\gamma\gamma}_t > 0.02\sqrt{s}$. There should be no charged
tracks matching any of the photon candidates.

\begin{figure}
  \begin{center}
\begin{tabular}{ll}
    \includegraphics[width=0.45\textwidth]{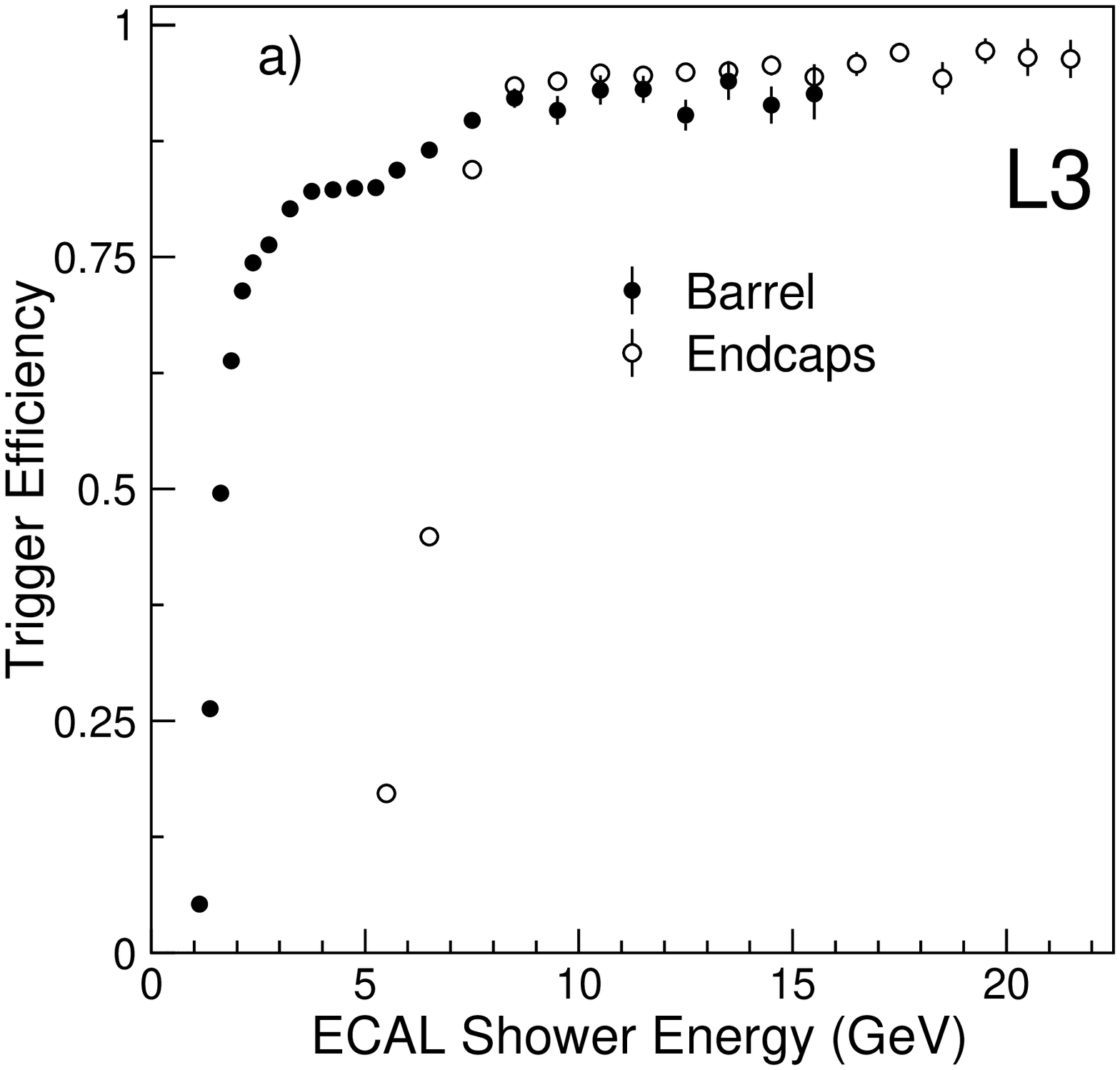} &
    \includegraphics[width=0.45\textwidth]{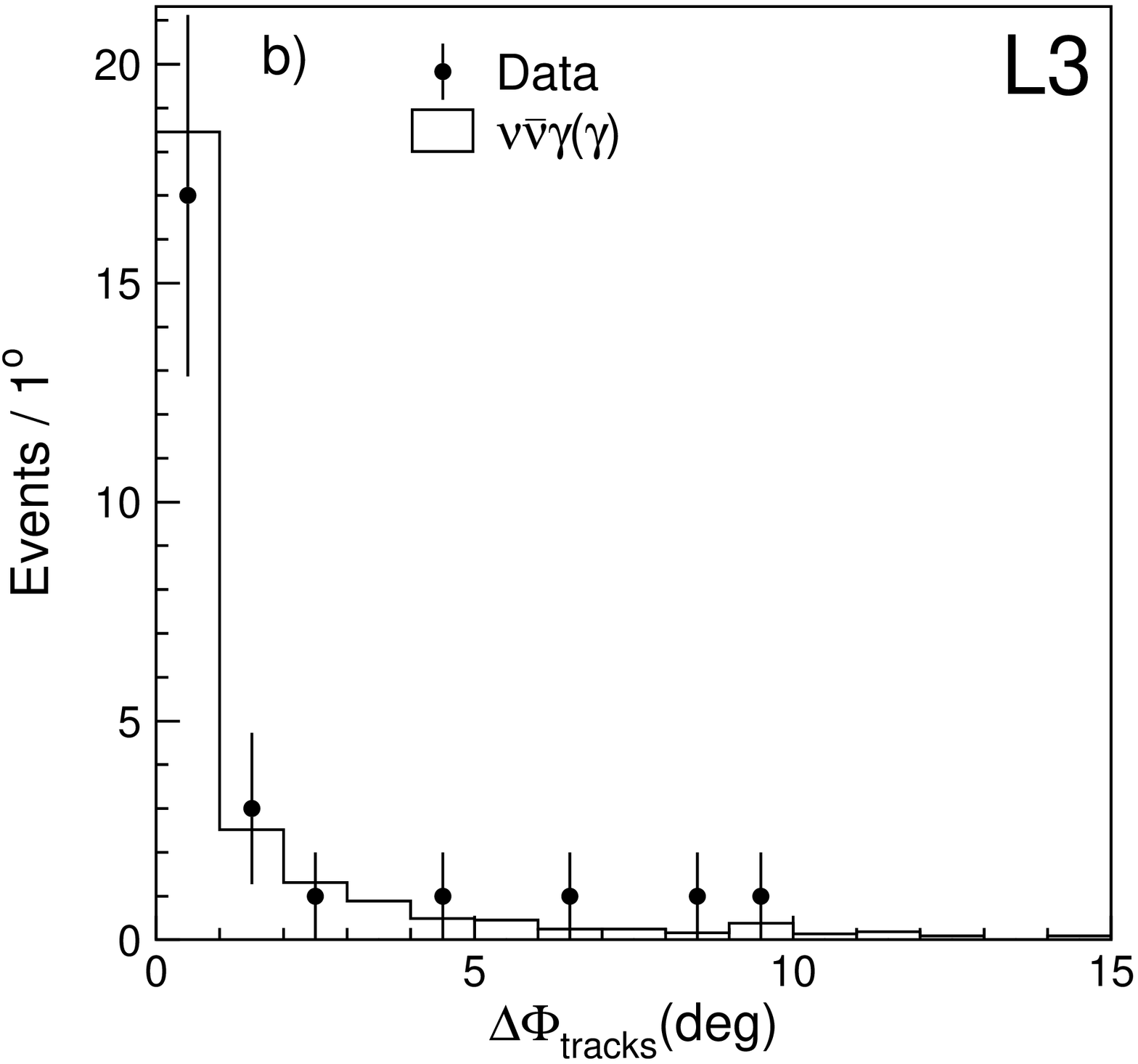} \\
    \includegraphics[width=0.45\textwidth]{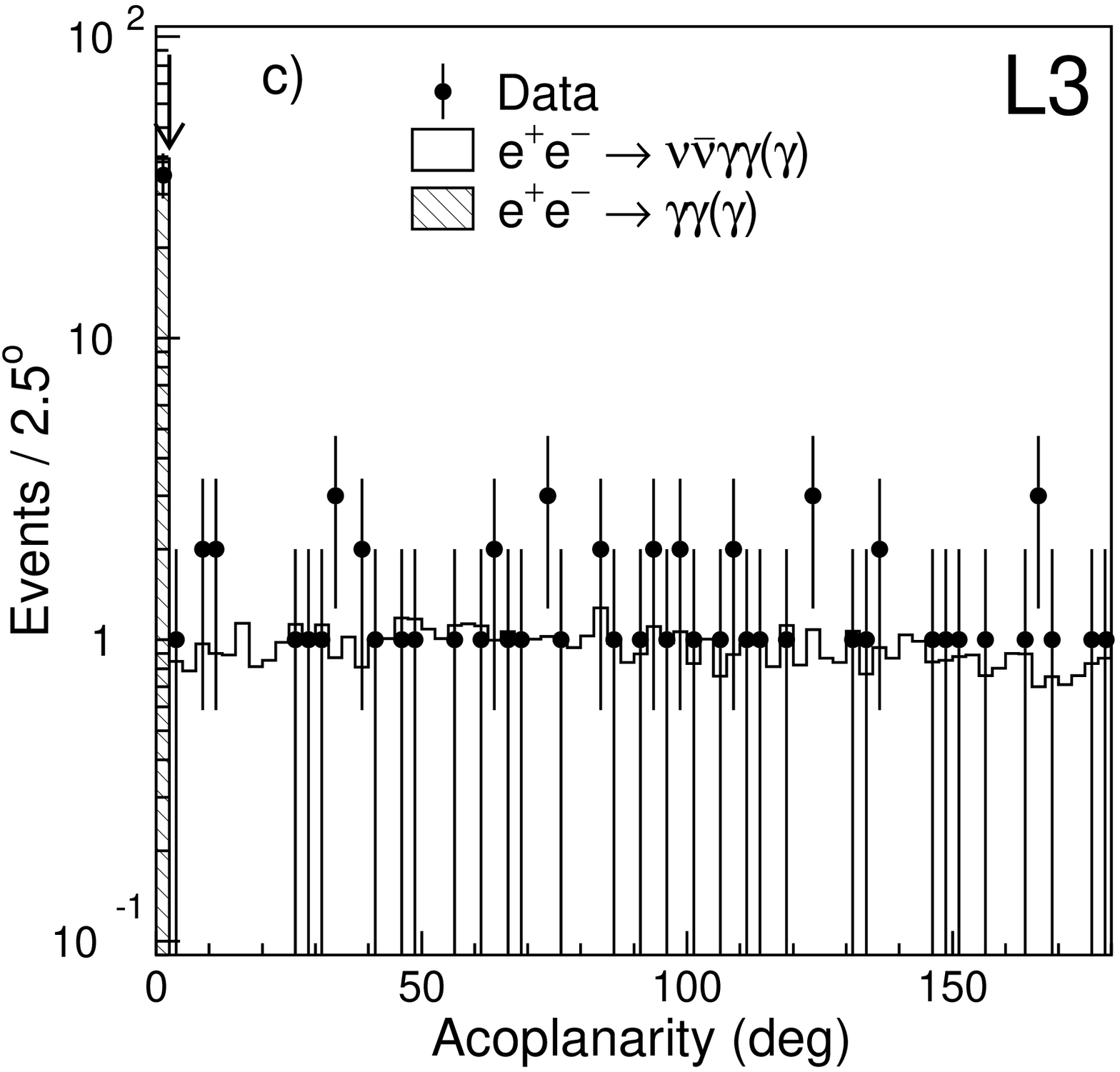} &
    \includegraphics[width=0.45\textwidth]{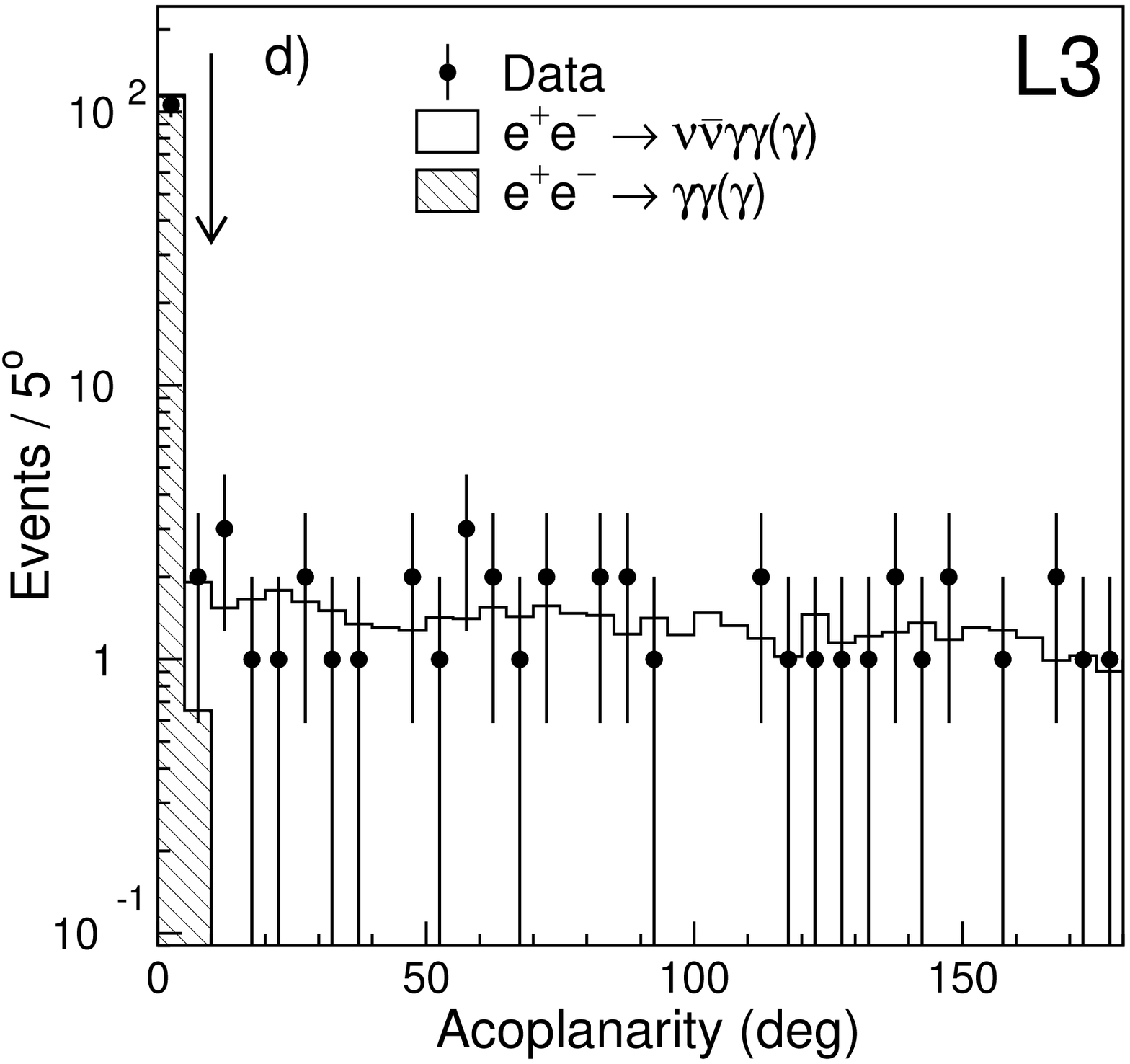} \\
\end{tabular}
    \icaption{\label{fig:sel_plots} a) Trigger efficiency as a
function of the ECAL shower energy.  Distributions of: b) the azimuthal
angle between two matching tracks for photons accepted by the conversion
selection in the barrel, c) the acoplanarity between the two most
energetic photons for ECAL showers which are not near the calorimeter edges and do
not contain dead channels, and d) for the case when at least one of
the showers does not satisfy these conditions.  The arrows indicate
the values of the cuts.  }
  \end{center}
\end{figure}

The acoplanarity between the two most energetic photons is required to
be greater than 2.5$^\circ$. About 20\% of the photon candidates are
either near the calorimeter edges or have a dead channel in the
$3\times 3$ matrix around the crystal with the maximum energy
deposition. For these events, the acoplanarity cut is relaxed to
10$^\circ$.  The distributions of the acoplanarity for events passing
all other selection cuts are shown in Figures~\ref{fig:sel_plots}c
and~\ref{fig:sel_plots}d.

In total, 101 multi-photon events are selected, with 114.8 expected
from the Standard Model processes. The purity of the selected sample
is 99.0\%, with the main background coming from the 
$\epem\rightarrow\gamma\gamma(\gamma)$ process.
Figures~\ref{fig:egamma}b and~\ref{fig:egamma}d show the distributions
of $M_{rec}$ and of the energy of the second most energetic photon,
$E_{\gamma_2}$.  Table~\ref{tab:sgamsel} gives the numbers of
multi-photon events selected at different values of $\sqrt{s}$
together with the Standard Model expectations.  The efficiencies of the
selection and the numbers of observed and expected events are given in
Table~\ref{tab:multiphoton} in bins of $M_{rec}$ and $E_{\gamma_2}$,
for the full sample and for the case in which both photons are in the
barrel.

%
%%%%%%%%%%%%%%%%%%%%%%%%%%%%%%%%%%%%%%%%%%%%%%%%%%%%%%%%%%%%%%%%%%%%%%%%%%%%%%
\subsection{Low Energy Single-Photon Selection}
%%%%%%%%%%%%%%%%%%%%%%%%%%%%%%%%%%%%%%%%%%%%%%%%%%%%%%%%%%%%%%%%%%%%%%%%%%%%%%
%

This selection extends the $P_t^\gamma$ range down to $0.008
\sqrt{s}$. It covers only the barrel region where a single-photon
trigger~\cite{sgtrig} is implemented with a threshold around 900~\mev,
as shown in Figure~\ref{fig:sel_plots}a.  In this region the
background due to radiative Bhabha scattering increases, requiring
additional cuts: no energy deposit is allowed in the forward
calorimeters, there must be no other ECAL cluster with energy greater
than 200~\mev, the energy in the hadron calorimeter must be less than
6~\gev\ and no tracks are allowed either in the TEC or in the muon
chambers. To further reduce  background from cosmic ray
 events not pointing to the
interaction region, cuts on the transverse shape of the photon shower
are also applied.

\begin{figure}
  \begin{center}
    \includegraphics[width=0.65\linewidth]{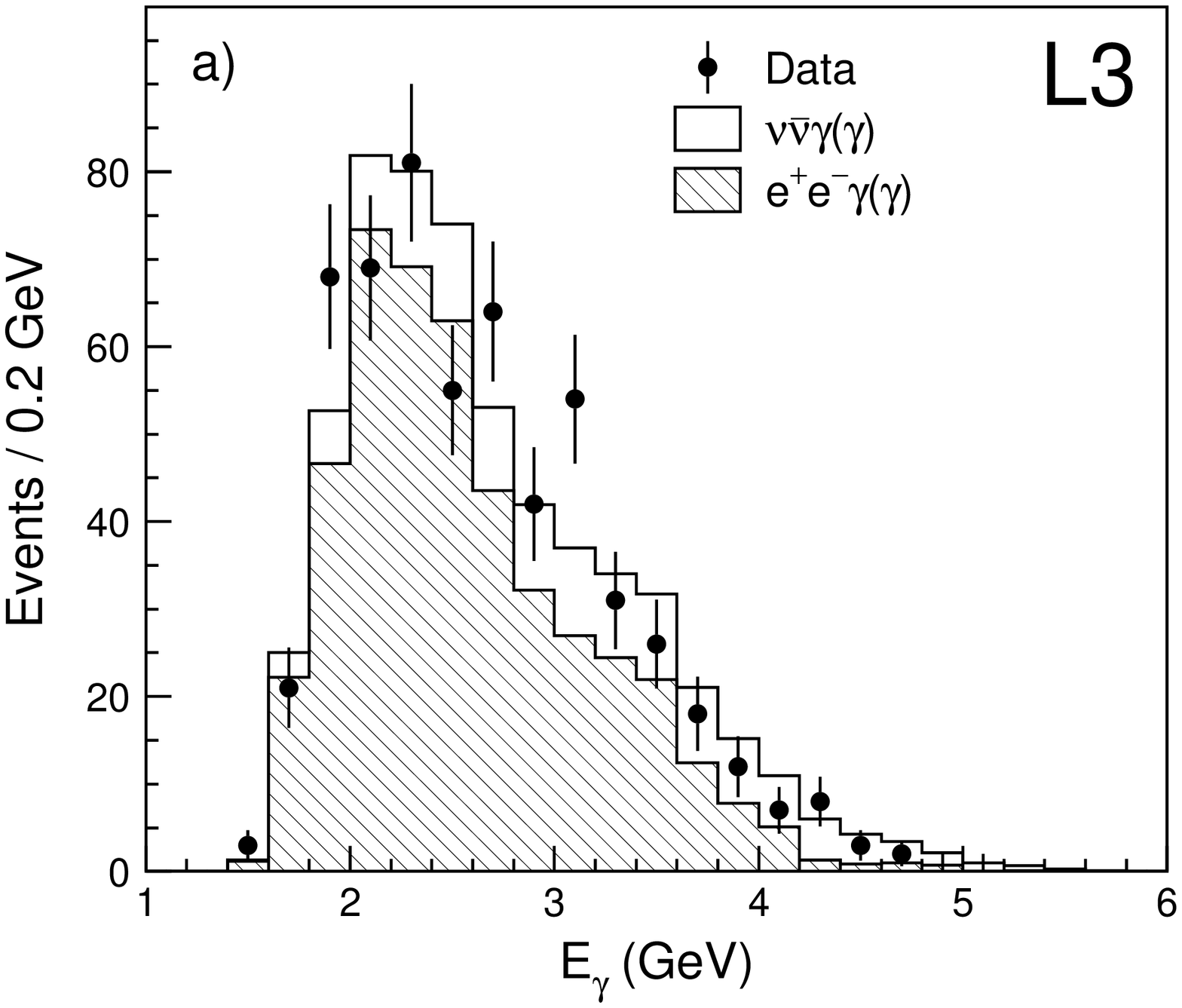} \\
 \includegraphics[width=0.65\linewidth]{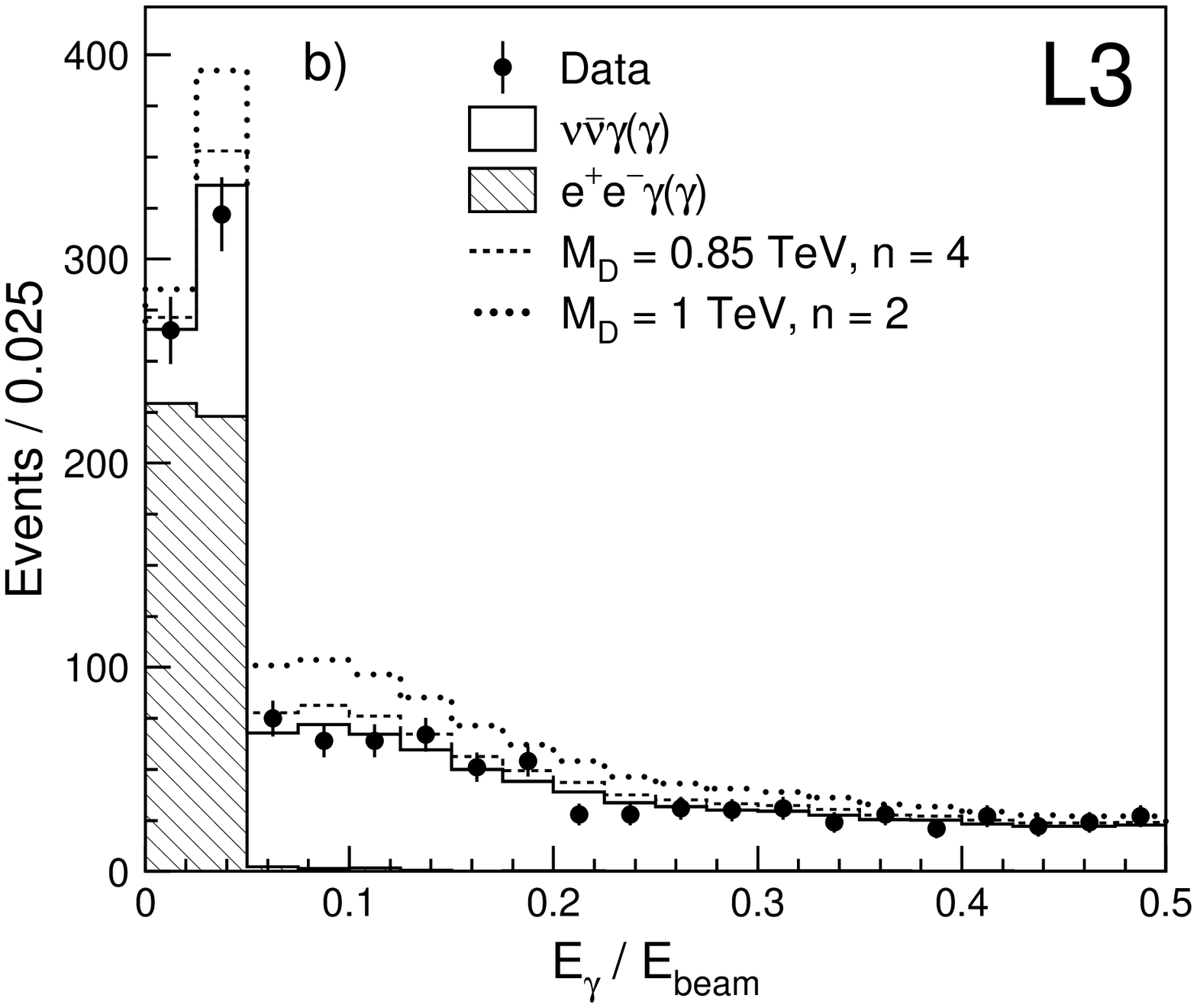}
  \end{center}
    \icaption{\label{fig:extrad_ggg}
Distributions of a) the photon energy for the low energy single-photon
selection and b) the ratio of the photon energy to the beam energy,
$x_\gamma$, for single-photon events from the combined high and low
energy single-photon selections. Signals for extra dimensions for $M_D
= 1~\mathrm{and}~0.85$~\TeV\ and $n = 2~\mathrm{and}~4$ are also
shown.
}
\end{figure}

The numbers of selected and expected events are listed in
Table~\ref{tab:sgamsel}.  In total, 566 events are selected in data
with an expectation of 577.8, where 124.2 events are expected from the
\epem\ \ra\ \nnbar\gam(\gam) process and 447.2 from the \epem\ \ra\
\epem\gam(\gam) process.  Figure~\ref{fig:extrad_ggg}a compares the
photon energy spectrum with the Monte Carlo predictions. The
normalisation of the \epem\ \ra\ \epem\gam(\gam) Monte Carlo is
verified with a data sample selected with less stringent selection
criteria.

Table~\ref{tab:low_sg} presents the numbers of  observed and
expected events, the efficiencies and the purities of the selected sample
in bins of $|\cos\theta_\gamma|$ and 
$x_{\gamma}=E_{\gamma}/E_{beam}$, where $E_{beam}$ is the
beam energy. Single-photon events with
$x_{\gamma} < 0.5$ from the combined high and low energy selections
are listed, and the corresponding $x_{\gamma}$\ distribution is shown
in Figure~\ref{fig:extrad_ggg}b.

%
%%%%%%%%%%%%%%%%%%%%%%%%%%%%%%%%%%%%%%%%%%%%%%%%%%%%%%%%%%%%%%%%%%%%%%%%%%%%%%
\section{Neutrino Production}
%%%%%%%%%%%%%%%%%%%%%%%%%%%%%%%%%%%%%%%%%%%%%%%%%%%%%%%%%%%%%%%%%%%%%%%%%%%%%%
%

The cross section of the process $\mathrm{e}^+\mathrm{e}^- \rightarrow
\nu \bar{\nu} \gamma (\gamma)$, where one or more
photons are observed,
 is measured in the kinematic region
$14^{\circ} < \theta_{\gamma} < 166^{\circ}$ and 
$P^{\gamma}_t > 0.02\sqrt{s}$  or $P^{\gamma\gamma}_t > 0.02\sqrt{s}$
using the high energy single-photon and the
multi-photon samples.  The average combined trigger and selection efficiency is estimated to be about 71\%
and is given in Table~\ref{tab:xsecs} as a function of $\sqrt{s}$
 together with the results of the cross section measurement
and the Standard Model expectations.

The systematic uncertainties on the cross section are listed in
Table~\ref{tab:syst}. The largest sources of systematics are the
uncertainty on the determinations of the trigger efficiency and of the
efficiency of the selection of converted photons, both due to the
statistics of control data samples. Equally large is the uncertainty
from Monte Carlo modelling, determined as the full difference between
the efficiencies obtained using the {\tt KKMC} and {\tt NUNUGPV} Monte
Carlo generators. Other uncertainties are due to the selection
procedure, assigned by varying the selection criteria, the Monte Carlo
statistics, the uncertainty on the measurement of the integrated
luminosity, the level of background from Standard Model processes and
cosmic rays and, finally, the accuracy of the ECAL calibration.  All
uncertainties, except that from Monte Carlo statistics, are fully
correlated over different values of $\sqrt{s}$.

\begin{figure}
  \begin{center}
    \includegraphics[width=0.9\textwidth]{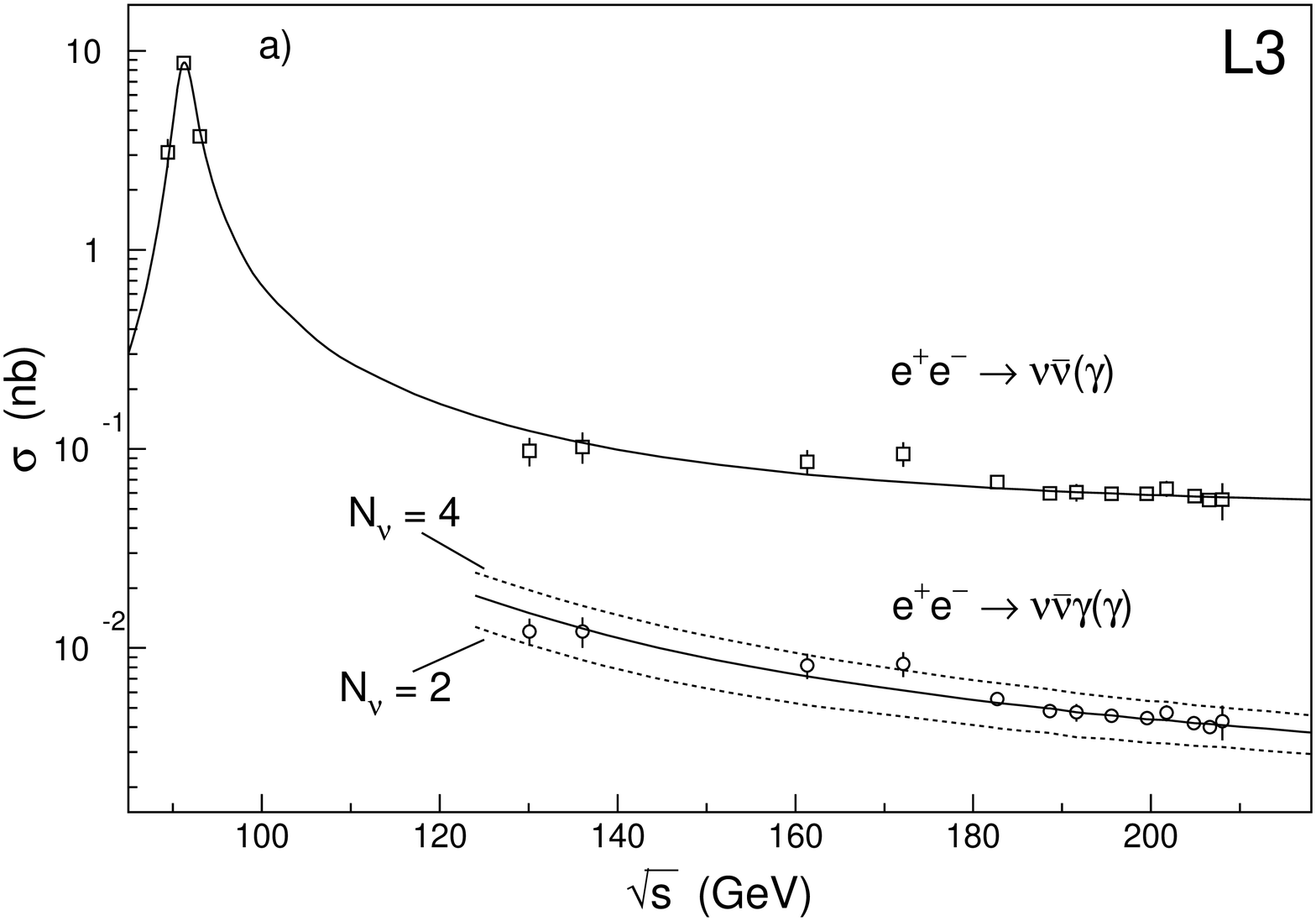}
    \includegraphics[width=0.9\textwidth]{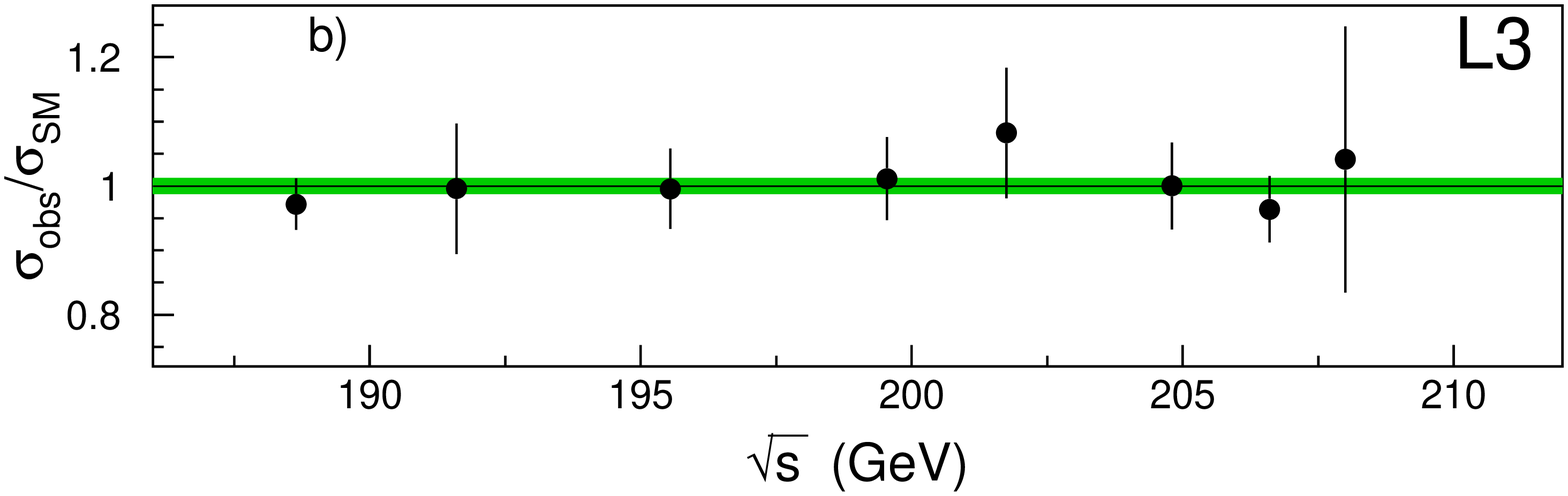}
  \end{center}
    \icaption{\label{fig:nnxsection} a) Cross sections of the \epem\
\ra\ \nnbar(\gam) and \epem\ \ra\ \nnbar\gam(\gam) processes as a
function of $\sqrt{s}$. 
The cross section of the latter process refers to the 
kinematic region defined in the text.
The full line represents the theoretical
prediction for $N_\nu=3$ and the dashed lines are predictions for
$N_\nu = 2$ and 4, as indicated. b) The ratio of the measured and the
Standard Model predicted cross sections as a function of $\sqrt{s}$. The shaded
region represents the theoretical uncertainty of 1\%~\cite{KK_theory}.
}
\end{figure}

Figure~\ref{fig:nnxsection} shows  the measured $\epem\ra\nnbar\gam(\gam)$
cross section  as a function
of $\sqrt{s}$, together with the Standard Model predictions and
measurements at lower $\sqrt{s}$~\cite{papgg99}.  The theoretical
uncertainty on the predicted cross section is 1\%~\cite{KK_theory}.
The extrapolation to the total cross section of the
$\epem\ra\nnbar(\gam)$ process, obtained using the {\tt KKMC} program,
is also shown in Figure~\ref{fig:nnxsection}.

\begin{figure}
  \begin{center}
    \includegraphics[width=0.9\textwidth]{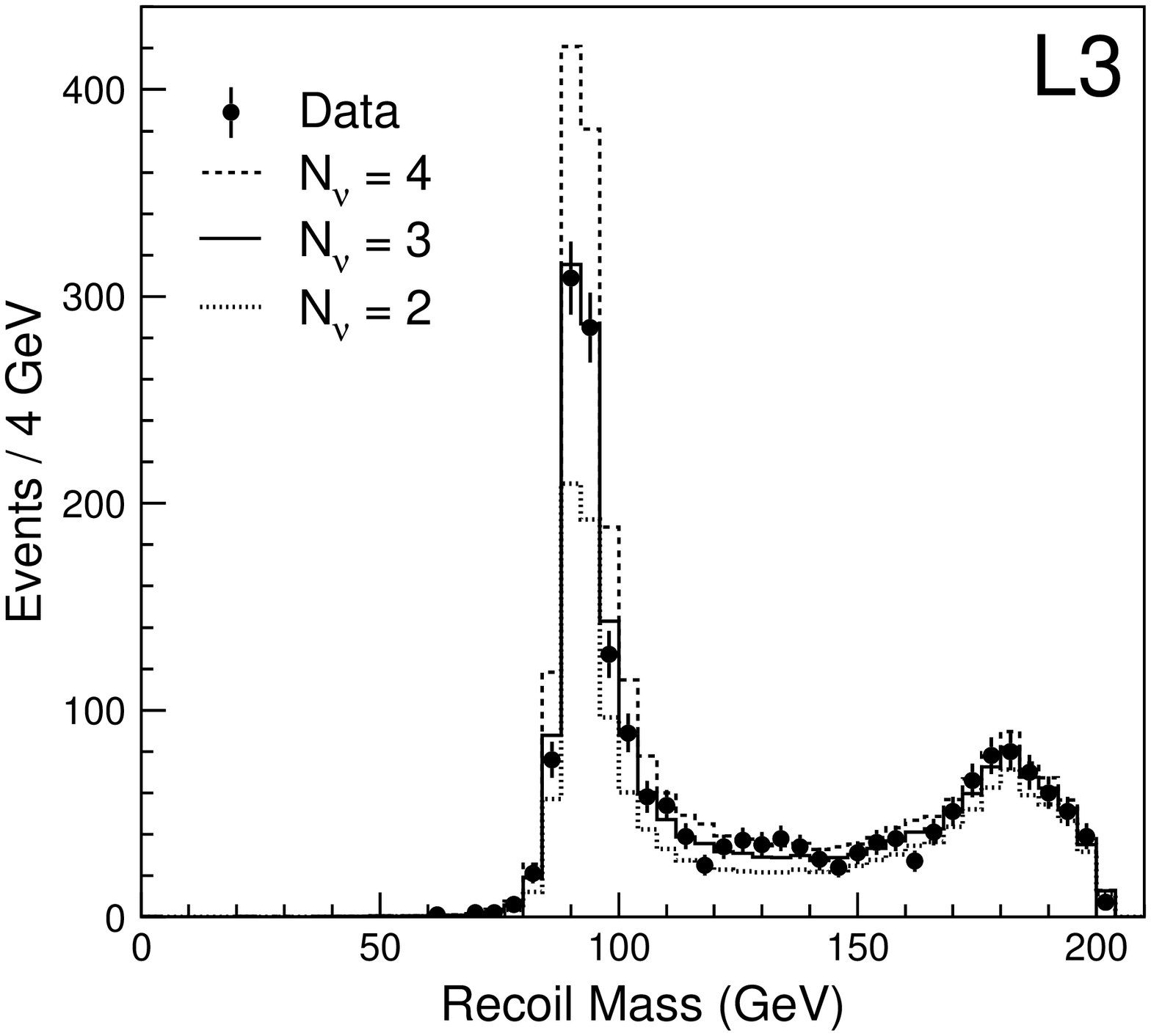}
  \end{center}
    \icaption{\label{fig:specspec} The recoil mass spectrum of the
 single- and multi-photon events compared to the expected spectra for $
 N_\nu = 2,3$ and 4.}
\end{figure}

To determine the number of light neutrino species, $N_\nu$, a binned
maximum likelihood fit is performed to the two dimensional
distribution of $M_{rec}$ {\it vs.} $|\cos\theta_\gam|$ for events
selected by the high energy single-photon and by the multi-photon
selections. The expectations for different values of $N_\nu$ are
obtained by a linear interpolation of the {\tt KKMC} predictions for
$N_\nu=2,3$ and 4. Due to the different contributions to the energy
spectrum from the $t$-channel $\nu_\e\bar\nu_\e$ production and the
$s$-channel \nnbar\ production, this method is more powerful than
using the total cross section measurement.  Figure~\ref{fig:specspec}
shows the $M_{rec}$ spectrum compared to the expectations for $ N_\nu
= 2, 3$ and 4. The result of the fit is:
 $$ N_\nu = 2.95 \pm 0.08 (stat) \pm 0.03 (syst) \pm 0.03(theory).$$
The systematic uncertainties are the same as for the cross section
measurement. 
The last uncertainty includes 
the theoretical uncertainty on the expected cross section~\cite{KK_theory}
as well as 
an additional uncertainty on the shape of the
recoil mass spectrum, estimated by comparing {\tt KKMC} with {\tt
NUNUGPV}.
  Combining this result with the L3 measurements
at $\sqrt{s}$ around the Z resonance~\cite{lep1nunug} and
above~\cite{papgg99}, gives
 $$ N_\nu = 2.98 \pm 0.05 (stat) \pm 0.04 (syst).$$
This result is in agreement with the Z lineshape
studies\cite{lep1lineshape}, while being sensitive to different
systematic and theoretical uncertainties.  It is more precise than the
present world average of measurements relying on the single-photon
method~\cite{hagiwara}.

%
%%%%%%%%%%%%%%%%%%%%%%%%%%%%%%%%%%%%%%%%%%%%%%%%%%%%%%%%%%%%%%%%%%%%%%%%%%%%%%
\section{Searches for New Physics}
%%%%%%%%%%%%%%%%%%%%%%%%%%%%%%%%%%%%%%%%%%%%%%%%%%%%%%%%%%%%%%%%%%%%%%%%%%%%%%
%

%
%%%%%%%%%%%%%%%%%%%%%%%%%%%%%%%%%%%%%%%%%%%%%%%%%%%%%%%%%%%%%%%%%%%%%%%%%%%%%%
\subsection{Extra Dimensions}
%%%%%%%%%%%%%%%%%%%%%%%%%%%%%%%%%%%%%%%%%%%%%%%%%%%%%%%%%%%%%%%%%%%%%%%%%%%%%%
%

Gravitons expected in theories with $n$ extra dimensions~\cite{qgrav1}
are produced via the \epem\ \ra\ \gam G process and are undetected,
giving rise to  a single photon and missing energy signature. This
reaction proceeds through $s$-channel photon exchange, $t$-channel
electron exchange and four-particle contact
interaction~\cite{extrad1}.

The efficiency for such a signal is derived in a $x_\gamma$ {\it vs.}
$|\cos{\theta_\gamma}|$ grid similar to that of Table~\ref{tab:low_sg}
and, together with the analytical differential cross
section~\cite{extrad1}, allows the calculation of the number of
expected signal events as a function of (1/$M_D$)$^{n+2}$, to which
the signal cross section is proportional.  Effects of ISR are taken
into account using the radiator function given in
Reference~\citen{berends}.  Since the photon energy spectrum from the
\epem\ \ra\ \gam G reaction is expected to be soft, only single-photon
events from the high and low energy samples with $x_\gamma <0.5$ are
considered. Effects of extra dimensions on the $x_\gamma$ distribution
are shown in Figure~\ref{fig:extrad_ggg}b.  The two-dimensional
distribution of $x_\gamma$ {\it vs.}  $| \cos\theta_\gamma |$ is fitted
including a term proportional to (1/$M_D$)$^{n+2}$ with the results
listed in Table~\ref{tab:extrad}.  While similar searches were
performed both at LEP~\cite{papgg99,paplepgg,l3lsg} and the
Tevatron~\cite{tevatron}, these results provide the most stringent
limits for $n<6$.

%
%%%%%%%%%%%%%%%%%%%%%%%%%%%%%%%%%%%%%%%%%%%%%%%%%%%%%%%%%%%%%%%%%%%%%%%%%%%%%%
\subsection{Model-Independent Searches}
%%%%%%%%%%%%%%%%%%%%%%%%%%%%%%%%%%%%%%%%%%%%%%%%%%%%%%%%%%%%%%%%%%%%%%%%%%%%%%
%

Single- and multi-photon events are used to investigate the \epem\ \ra\
XY and \epem\ \ra\ XX processes where X and Y are massive neutral undetectable
particles and the X \ra\ Y\gam\ decay occurs with a 100\% branching
ratio. Flat photon energy and polar angle distributions are assumed.

\begin{figure}
  \begin{center}
    \begin{tabular}{ll}
      \includegraphics[width=0.45\textwidth ]{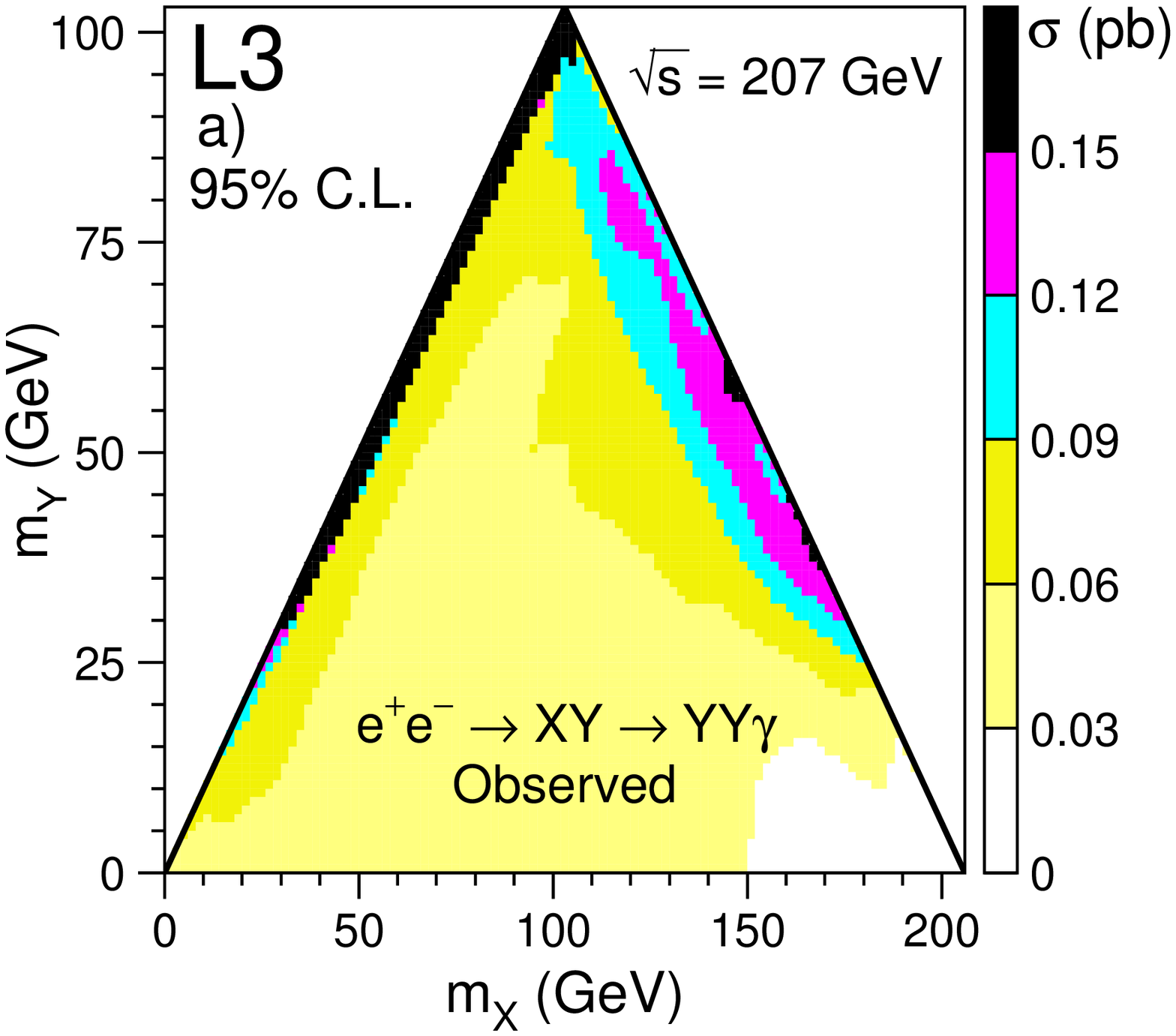}& 
      \includegraphics[width=0.45\textwidth ]{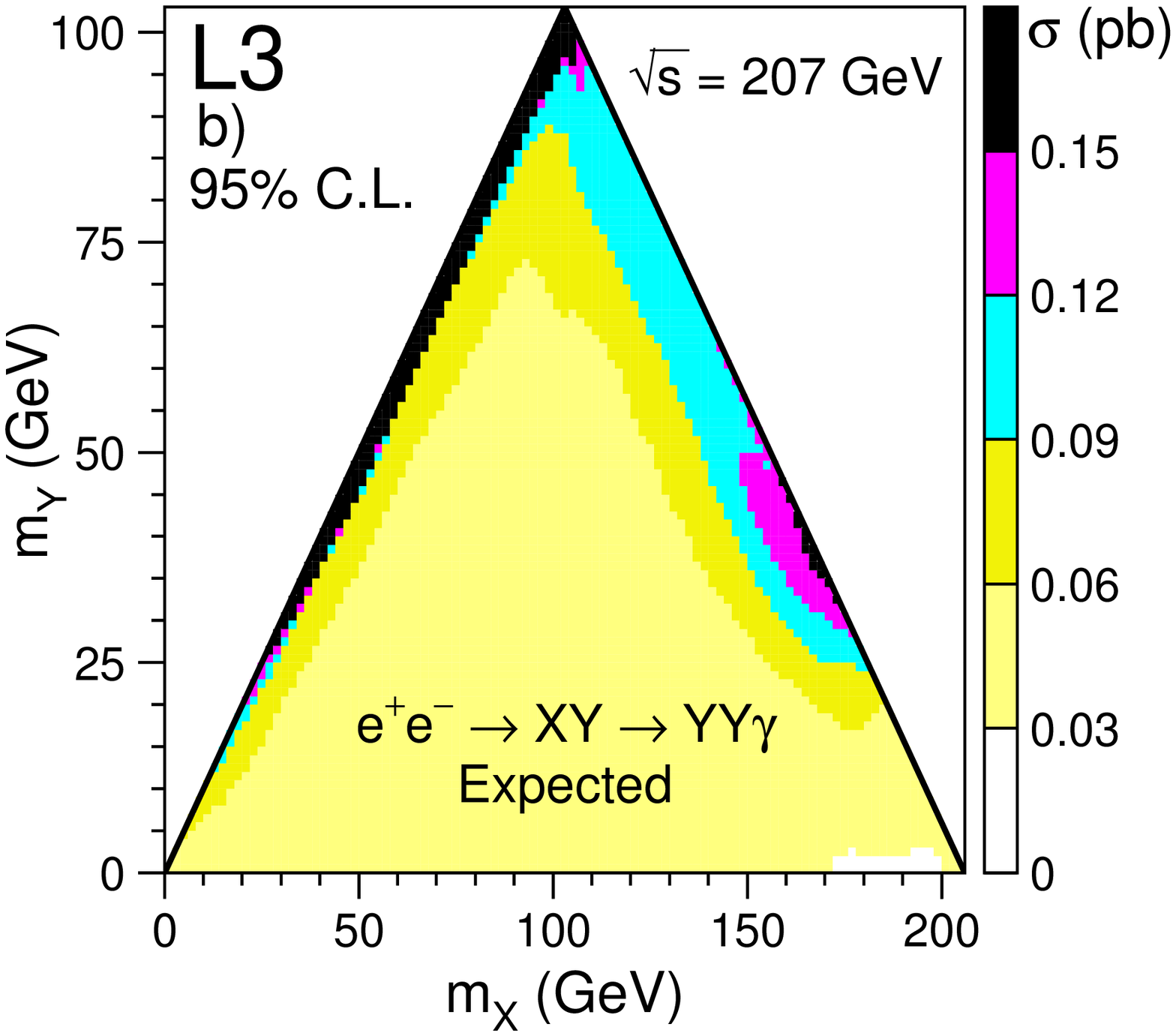}\\
      \includegraphics[width=0.45\textwidth ]{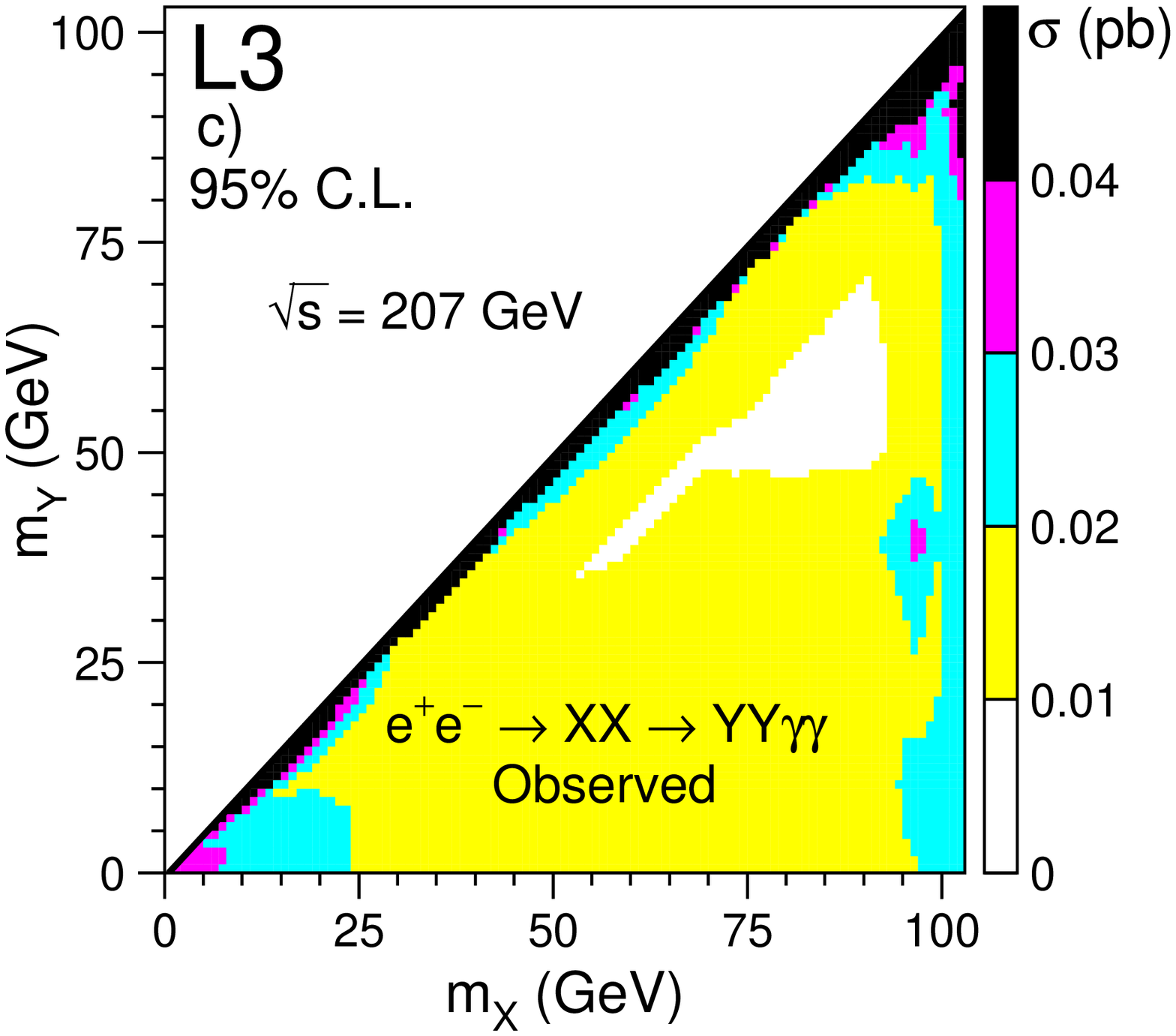}& 
      \includegraphics[width=0.45\textwidth ]{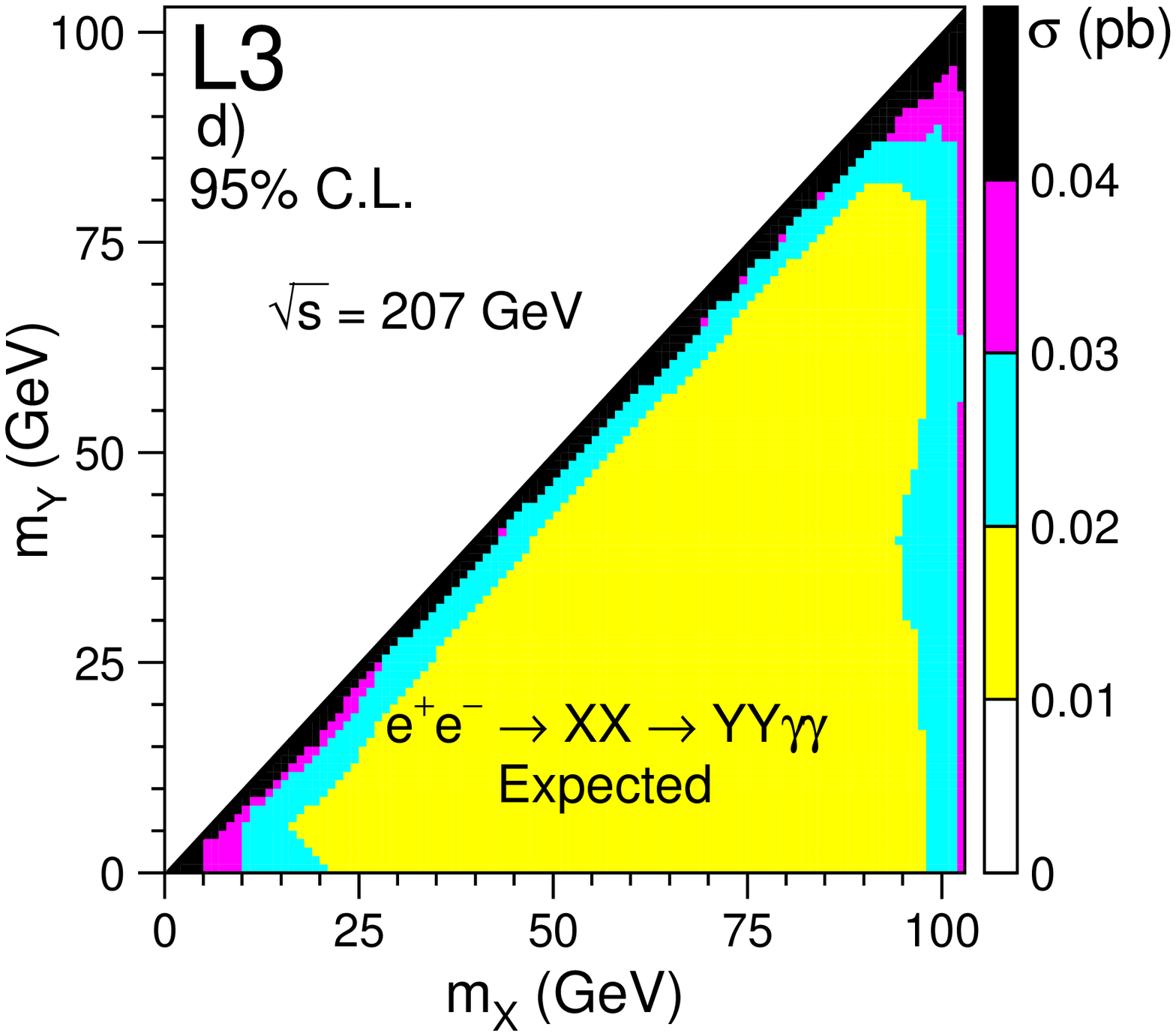}\\
    \end{tabular}
  \end{center}
  \icaption{\label{fig:n1ne2}Cross section upper limits at 95\%
confidence level for model-independent searches: a) observed and b)
expected for the process \ee \ra\ XY \ra\ YY\gam\ and c) observed and
d) expected for the process \ee \ra\ XX \ra\ YY\gam\gam.  The limits
are obtained for $\rts = 207 \GeV$. Data collected at lower $\rts$ are
included assuming the signal cross sections to scale as $\beta_0/s$, where
$\beta_0$ is defined in the text.}
\end{figure}

For the \epem\ \ra\ XY search, a fit is performed to the $M_{rec}$
distribution, whereas for the \epem\ \ra\ XX channel, a discriminant
variable is built~\cite{papgg99} which includes $M_{rec}$, the
energies of the two most energetic photons, their polar angles and the
polar angle of the missing momentum vector.  No deviation from the
Standard Model expectations is observed and cross section limits are
derived for all allowed values of the masses $m_X$ and $m_Y$, in steps
of 1~\GeV.  The observed and expected limits are shown in
Figure~\ref{fig:n1ne2} in the $m_Y$ {\it vs.} $m_X$ plane.  The limits
are obtained at $\rts = 207 \GeV$, data collected at lower $\rts$ are
included assuming the signal cross section to scale as ${\beta}_0/s$,
where ${\beta}_0 = \sqrt{1-2(x_1+x_2)+(x_1-x_2)^2}$ with $x_1 =
m_X^2/s$ and $x_2 = m_X^2/s$ or $x_2 = m_Y^2/s$ for the \epem\ \ra\ XX
and \epem\ \ra\ XY searches, respectively\footnote{We assume that the
matrix elements of both processes do not depend on $\sqrt{s}$.}.

%
%%%%%%%%%%%%%%%%%%%%%%%%%%%%%%%%%%%%%%%%%%%%%%%%%%%%%%%%%%%%%%%%%%%%%%%%%%%%%%
\subsection{ Neutralino Production in SUGRA Models}
%%%%%%%%%%%%%%%%%%%%%%%%%%%%%%%%%%%%%%%%%%%%%%%%%%%%%%%%%%%%%%%%%%%%%%%%%%%%%%
%

In gravity-mediated SUSY breaking models (SUGRA) the gravitino is
heavy (\( 100 \gev \lesssim \MG \lesssim 1 \tev \)) and does not play
a role in the production and decay of SUSY particles. The lightest
neutralino is the lightest supersymmetric particle (LSP), which is
stable under the assumption of R-parity~\cite{rparity} conservation
and escapes detection due to its weakly interacting nature. In this
scenario, single- or multi-photon signatures arise from neutralino
production through the processes \epem\ra\ \chinon\chinonn\ and
\epem\ra\ \chinonn\chinonn\ followed by the decay
\chinonn\ \ra\ \chinon\gam~\cite{neuprod}. The signal topologies are
 similar to the ones assumed in the model-independent searches 
described above,
and comparable cross section limits are derived.

\begin{figure}
  \begin{center}
    \includegraphics[width=0.9\textwidth]{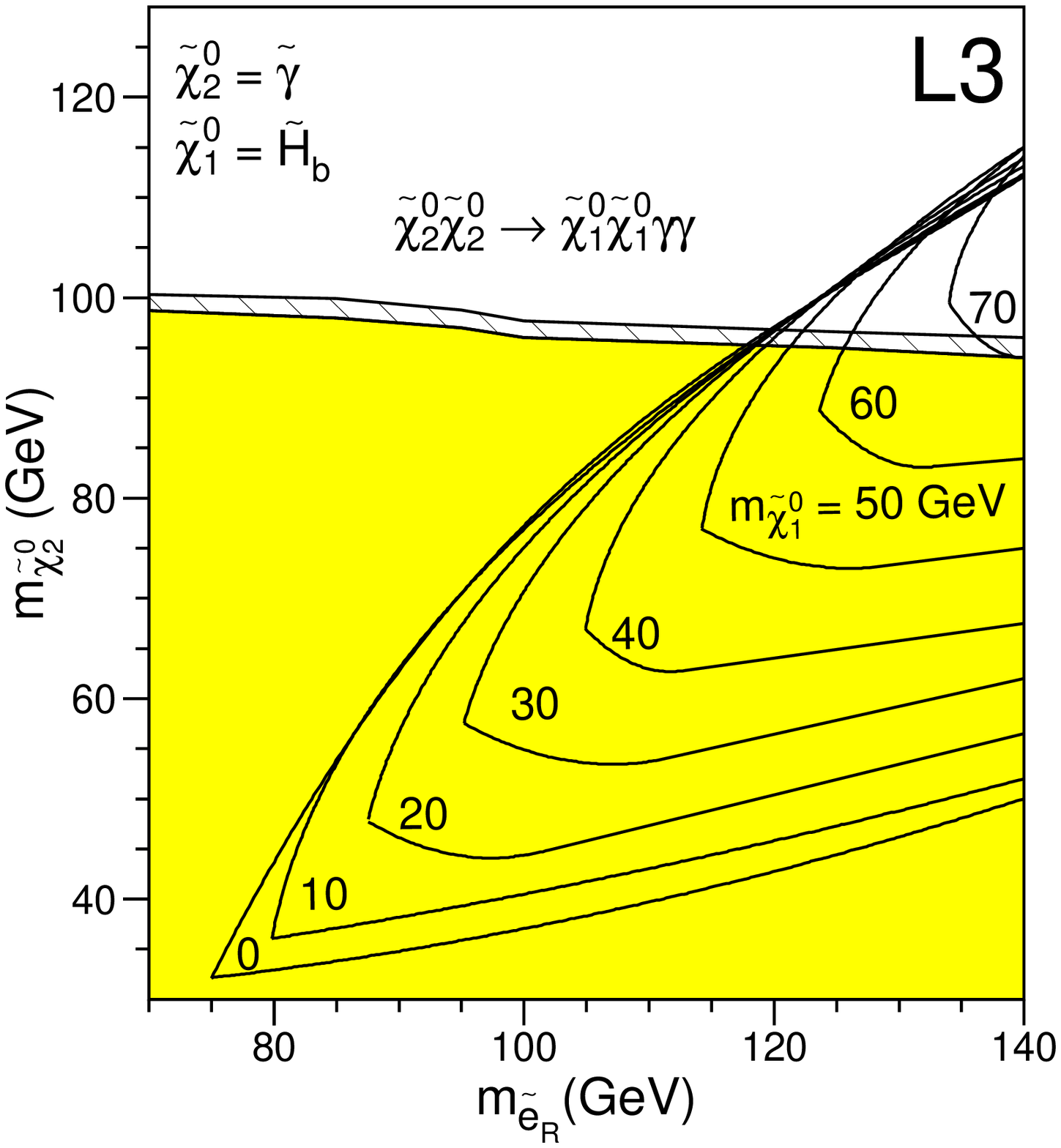}
  \end{center}
  \icaption{\label{fig:n2ne2}Region excluded at 95\% confidence level
  in the $\Mchii$ {\it vs.} $\Mser$ plane. The shaded region
  corresponds to \Msel\ $\gg$ \Mser\ and the hatched region is
  additionally excluded when \Msel\ = \Mser.  The mass difference
  between \chinonn\ and \chinon\ is assumed to be greater than
  10~\GeV.  Regions kinematically allowed for the CDF event
  \protect\cite{cdfinterp12} as a function of \Mchi\ are also indicated.}
\end{figure}

The one-loop \chinonn\ \ra\ \chinon\gam\ decay has a branching
fraction close to 100\% if one of the two neutralinos is pure photino
and the other pure higgsino~\cite{neugam}.  This scenario is suggested
by an interpretation~\cite{cdfinterp12} of the rare \e\e\gam\gam\
event observed by CDF~\cite{cdfevent2}.  With this assumption, and
using the results of the search for the \epem\ra\ \chinonn\chinonn\
process, a lower limit on the \chinonn\ mass is calculated as a
function of the right-handed scalar electron mass, \Mser , using the
most conservative cross section upper limit for any mass difference
between \chinonn\ and \chinon\ greater than 10~\GeV. Two distinct
scenarios are investigated: $\Msel=\Mser$ and $\Msel \gg \Mser$, where
\Msel\ is the mass of the left-handed scalar electron.
Figure~\ref{fig:n2ne2} shows the excluded region in the \Mchii\ {\it
vs.} \Mser\ plane.  The regions kinematically allowed from a study of
the CDF event~\cite{cdfinterp12} are also indicated.

%
%%%%%%%%%%%%%%%%%%%%%%%%%%%%%%%%%%%%%%%%%%%%%%%%%%%%%%%%%%%%%%%%%%%%%%%%%%
\subsection{Superlight Gravitinos}
%%%%%%%%%%%%%%%%%%%%%%%%%%%%%%%%%%%%%%%%%%%%%%%%%%%%%%%%%%%%%%%%%%%%%%%%%%
%

When the scale of local supersymmetry breaking is decoupled from the
breaking of global supersymmetry, as in no-scale supergravity models
\cite{noscalesugra}, the gravitino becomes ``superlight'' (\(
10^{-6}\ev \lesssim \MG \lesssim 10^{-4}\ev \)) and is produced not
only in SUSY particle decays but also directly, either in
pairs~\cite{zwirner} or associated with a
neutralino~\cite{lnz}. Pair-production of gravitinos with ISR,
$\epem\ra\gravin\gravin\gamma$, leads to a single-photon signature
which also arises from the \epem\ra\ \gravin\chinon\ process with
$\chinon\ra\gravin\gamma$.

If the mass of the next-to-lightest supersymmetric particle (NLSP) is
greater than $\sqrt{s}$, the process $\epem\ra\gravin\gravin\gamma$ is
the only reaction to produce SUSY particles. Its properties are
similar to those of extra dimensions signals and its cross section is
proportional to 1/$\MG^4$. A two-dimensional fit to the $x_\gamma$
{\it vs.}  $| \cos\theta_\gamma |$ distribution gives:
$$
  \MG > 1.35\times 10^{-5}\ev,
$$
at 95\% confidence level, corresponding to a lower limit on the SUSY
breaking scale $\sqrt{F} > 238 \gev$.  The expected lower limit on the
gravitino mass is $1.32\times 10^{-5}\ev$.

The reaction \epem\ \ra\ \gravin\chinon\ proceeds through $s$-channel
Z exchange and $t$-channel \susylr{\e} exchange. Efficiencies for this
process range between 68\% for \Mchi\ = 0.5~\gev{} and 75\% 
at the kinematic limit. Cross section upper limits are derived at
$\rts = 207 \GeV$ from the photon energy spectrum and are shown in
Figure~\ref{fig:n1gra}a.  Data collected at lower $\rts$ are included
assuming the signal cross section to scale as $\beta^8$~\cite{lnz}, where
$\beta$ is the neutralino relativistic velocity.
% = \sqrt{1-\Mchi^2/s}$~\cite{lnz}.

The no-scale SUGRA LNZ model~\cite{lnz} has only two free parameters,
\MG\ and \Mchi, and considers the neutralino to be almost pure bino
and to be the NLSP. Its dominant decay channel is \chinon\ \ra\
\gravin\gam, and a  contribution from the decay into Z for \( \Mchi
\gtrsim 100 \gev \) is taken into account.
 Figure~\ref{fig:n1gra}c shows the excluded
regions in the \MG\ {\it vs.} \Mchi\ plane. Gravitino masses below
$10^{-5} \eV$ are excluded for neutralino masses below 172~\GeV.

%
%%%%%%%%%%%%%%%%%%%%%%%%%%%%%%%%%%%%%%%%%%%%%%%%%%%%%%%%%%%%%%%%%%%%%%%%%%
\subsection{The {\boldmath \epem\ \ra\ \chinon\chinon\ \ra\ 
\gravin\gam\gravin\gam\ } Process in GMSB Models}
%%%%%%%%%%%%%%%%%%%%%%%%%%%%%%%%%%%%%%%%%%%%%%%%%%%%%%%%%%%%%%%%%%%%%%%%%%
%

In models with gauge-mediated SUSY breaking (GMSB) \cite{gmsb}, a
light gravitino (\( 10^{-2}\ev \lesssim \MG \lesssim 10^2\ev \)) is
the LSP.  If the lightest neutralino is the NLSP, it decays
predominantly through \chinon\ \ra\ \gravin\gam, and pair-production
of the lightest neutralino leads to a two-photon plus missing energy
signature. The selection described in this Letter is devised for photons
originating from the interaction point, and  the following limits
are derived under the assumption of a
neutralino mean decay length shorter than 1~cm.

\begin{figure}
  \begin{center}
    \begin{tabular}{ll}
      \includegraphics[width=0.45\linewidth]{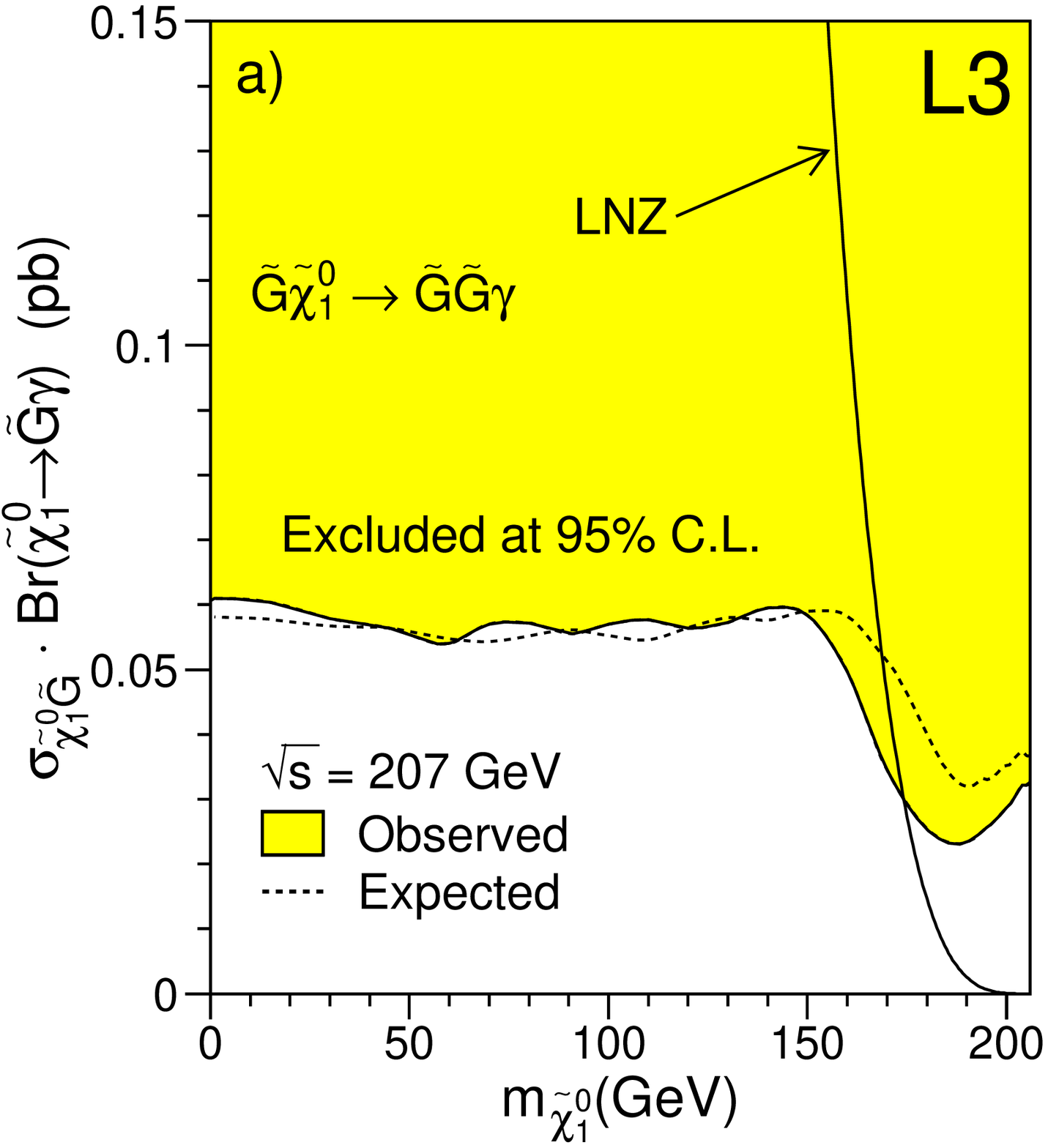} &
      \includegraphics[width=0.45\linewidth]{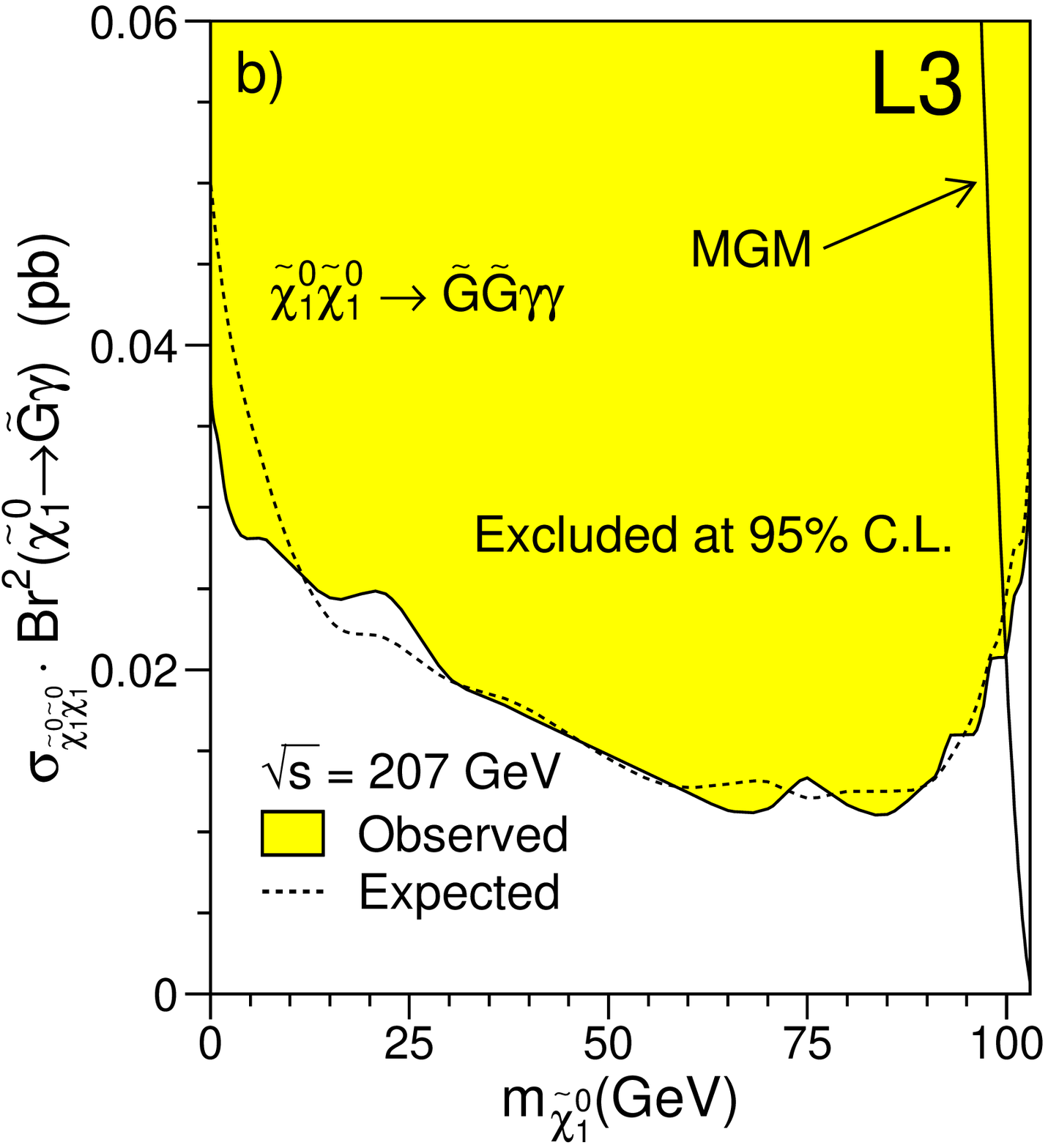} \\
      \includegraphics[width=0.45\linewidth]{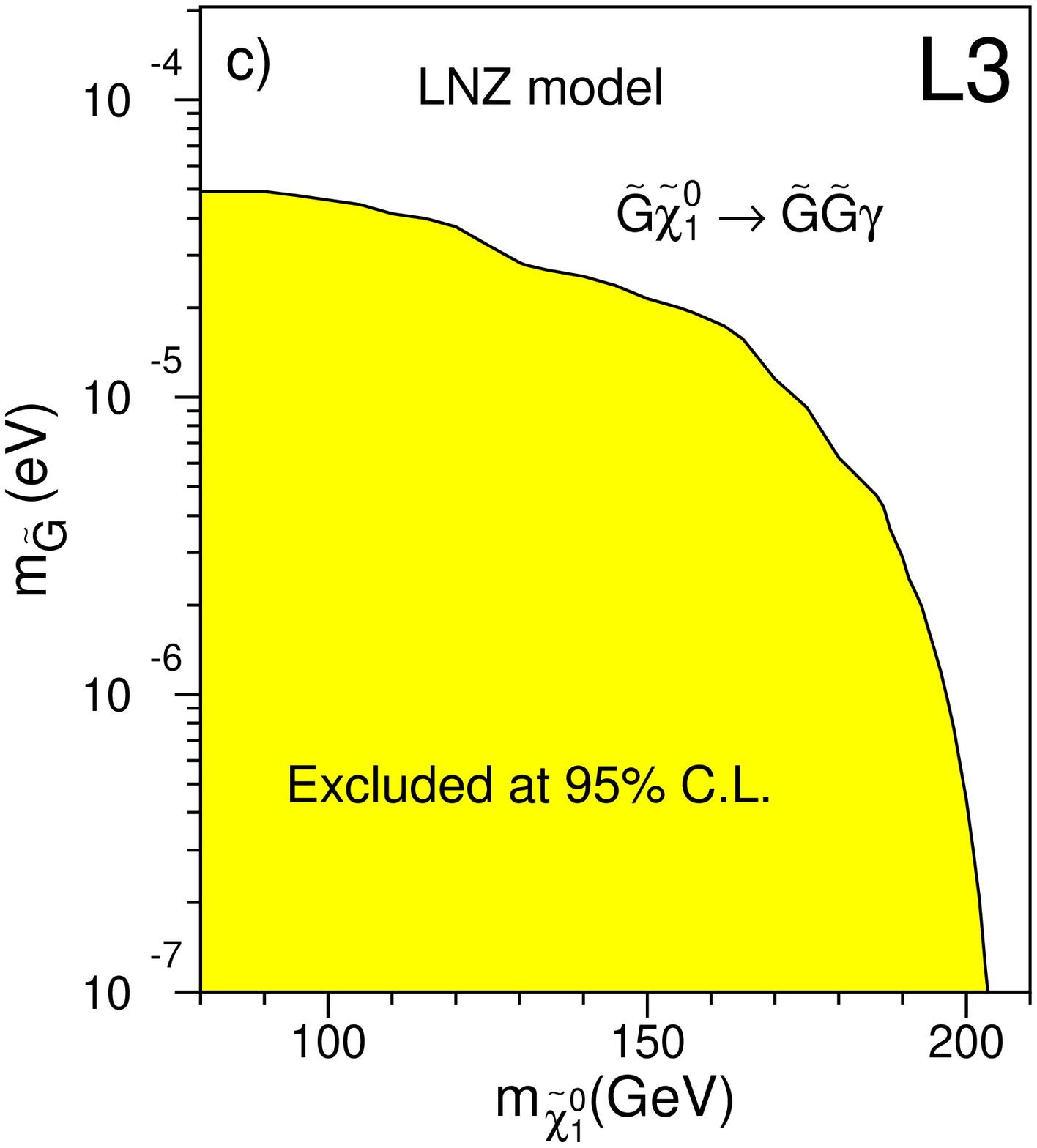} &
      \includegraphics[width=0.45\linewidth]{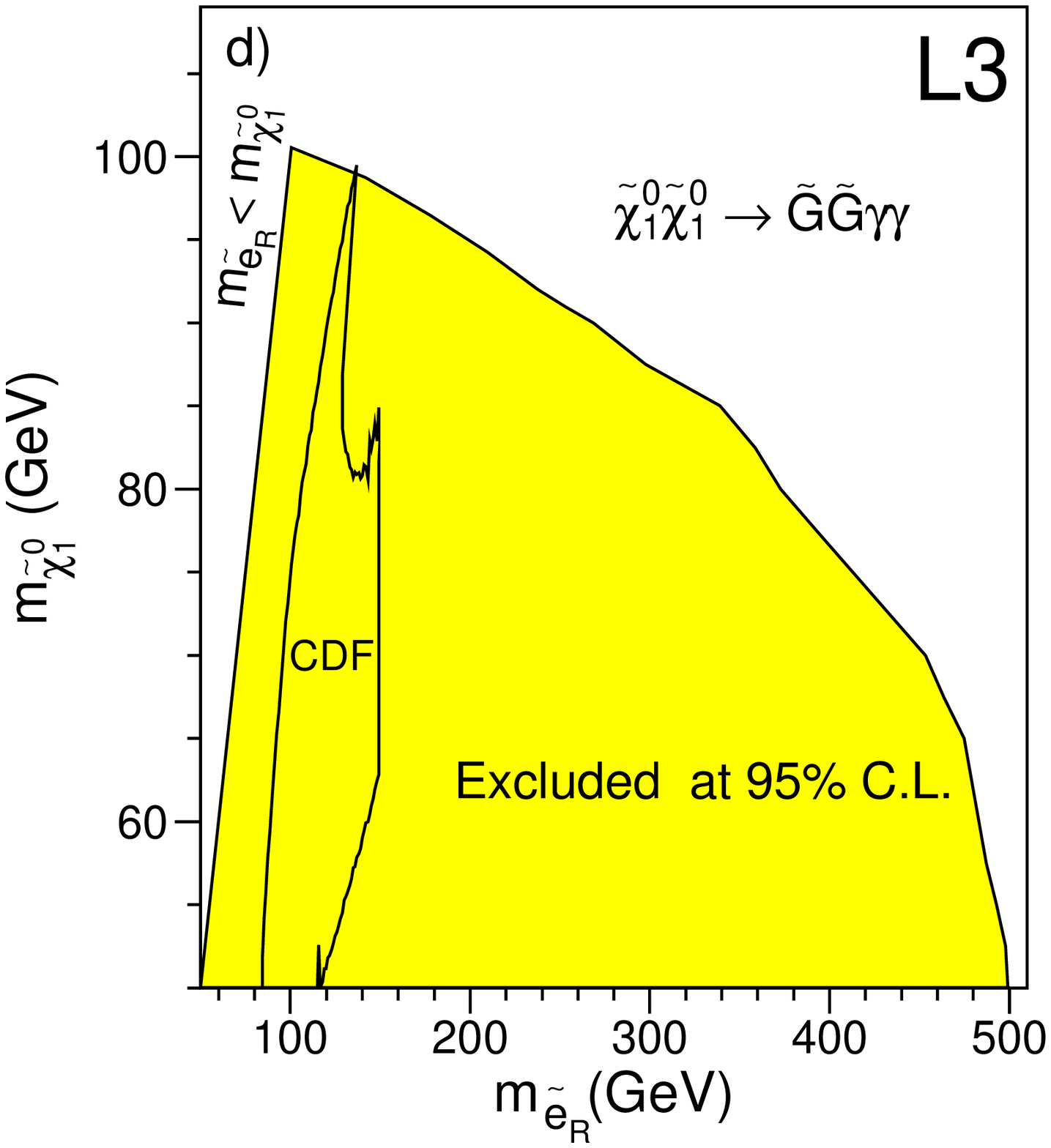} \\
    \end{tabular}
  \end{center}
  \icaption{\label{fig:n1gra} Observed and expected 95\% confidence
    level upper limits at $\rts = 207 \GeV$ on the production cross
    section for the processes a) \epem\ \ra\ \gravin\chinon\ \ra\
    \gravin\gravin\gam\ and b) \epem\ \ra\ \chinon\chinon\ \ra\
    \gravin\gravin\gam\gam.  The cross section predicted by the
    LNZ model~\cite{lnz} for $\MG = 10^{-5} \eV$ is also shown
    in a), while the prediction of the MGM model  is shown
    in b).  Regions excluded for c) the LNZ model in the \MG\ {\it vs.}
    \Mchi\ plane, and for d) a pure bino neutralino
    model in the \Mchi\ {\it vs.} \Mser\
    plane. The interpretation of the CDF event in the scalar
    electron scenario \protect\cite{ln96} is also shown in~d).}
\end{figure}

The same discriminant variable as in the \epem\ \ra\ XX \ra\ YY\gam\gam\
search is used and signal efficiencies are obtained which vary between
35\% for $\Mchi = 0.5 \gev$ and 70\% for $\Mchi \gtrsim 100\gev$.  No
deviations from the Standard Model are observed and upper limits on
the cross section are derived as a function of \Mchi\ at $\rts = 207
\GeV$, as displayed in Figure~\ref{fig:n1gra}b. 
Data collected at
lower $\rts$ are included assuming the signal cross section to scale
according to the MGM  model~\cite{MGM}.
The signal cross section
predicted by the MGM model is also shown in Figure~\ref{fig:n1gra}b.
In this model, the neutralino is  pure bino,
and $\Msel=1.1\times\Mchi$ and $\Mser=2.5\times\Mchi$.  
A 95\% confidence level limit on the neutralino mass is obtained as:
 $$ \Mchi > 99.5 \gev.$$ 
Figure~\ref{fig:n1gra}d shows the exclusion
region in the \Mchi\ {\it vs.} \Mser\ plane obtained 
after relaxing the mass relations of the MGM.
The region
suggested by an interpretation~\cite{ln96}
 of the \e\e\gam\gam\ event observed by
CDF is also shown. This interpretation is ruled out by this analysis.

%%%%%%%%%%%%%%%%%%%%%%%%%%%%%%%%%%%%%%%%%%%%%%%%%%%%%%%%%%%%%%5
\section{Conclusions}
%%%%%%%%%%%%%%%%%%%%%%%%%%%%%%%%%%%%%%%%%%%%%%%%%%%%%%%%%%%%

The high performance BGO calorimeter and the dedicated triggers of the L3
detector are used to select events with one or more photons and missing
energy in the high luminosity and centre-of-mass energy data sample
collected at LEP. Single- and multi-photon events
 with transverse momentum as low as $0.008\sqrt{s}$
are considered. The numbers of selected events agree with the expectations
from Standard Model processes and are given as a function of different
phase space variables in the form of tables which can be used to test future
models. The cross section for the process 
$\mathrm{e}^+\mathrm{e}^- \rightarrow \nu \bar{\nu} \gamma (\gamma) $
 is measured with high
precision as a function of $\sqrt{s}$, and is found to be
in agreement with the Standard Model prediction. From these and lower
energy data, the most precise direct determination of the number of light
neutrino families is derived as:
$$N_\nu=2.98 \pm 0.05(stat) \pm 0.04 (syst).$$
Model independent searches for the production of new invisible massive
particles in association with photons do not reveal any deviations
from the Standard Model expectations
 and
upper limits on the production cross sections are derived. 
Severe constraints are
placed on models with large extra dimensions and several SUSY scenarios,
excluding their manifestations at LEP.

%
%%%%%%%%%%%%%%%%%%%%%%%%%%%%%%%%%%%%%%%%%%%%%%%%%%%%%%%%%%%%%%%%%%%%%%%%%%%%%%%
% Bibliography
%%%%%%%%%%%%%%%%%%%%%%%%%%%%%%%%%%%%%%%%%%%%%%%%%%%%%%%%%%%%%%%%%%%%%%%%%%%%%%
%
% Style file to use with mcite.
% Use l3style with just cite.
\bibliographystyle{l3style}
\bibliography{mybibnew}

\newpage
\typeout{   }     
\typeout{Using author list for paper 279 -  }
\typeout{$Modified: Jul 15 2001 by smele $}
\typeout{!!!!  This should only be used with document option a4p!!!!}
\typeout{   }
%
%
%
%  L A T E X  version!!
%
%
% Make sure that the Lep package has been used!
%\input{Lep.sty}%
%
%\ifx\LepCalled\undefined%
%\typeout{     }%
%\typeout{!!!!!!!!!!!!!!!!!!!!!!!!!!!!!!!!!!!!!!!!!!!!!!!!!!!!!!!!!!!}%
%\typeout{Yikes.  You haven't used the Lep package!}%
%\typeout{Please put \protect\usepackage\protect{Lep\protect} in your preamble,
%         followed by}%
%\typeout{\protect\Lep\protect{1\protect} or \protect\Lep\protect{2\protect}}%
%\typeout{     }%
%\typeout{For now you will get a Lep phase 2 authorlist (may not be right!).}%
%\typeout{!!!!!!!!!!!!!!!!!!!!!!!!!!!!!!!!!!!!!!!!!!!!!!!!!!!!!!!!!!!}%
%\typeout{     }%
%\Lep{2}\fi%

\newcount\tutecount  \tutecount=0
\def\tutenum#1{\global\advance\tutecount by 1 \xdef#1{\the\tutecount}}
\def\tute#1{$^{#1}$}
\tutenum\aachen            % 1
\tutenum\nikhef            % 2
\tutenum\mich              % 3
\tutenum\lapp              % 4
\tutenum\basel             % 5
\tutenum\lsu               % 6
\tutenum\beijing           % 7
\tutenum\bologna           % 8
\tutenum\tata              % 9 
\tutenum\ne                % 10
\tutenum\bucharest         % 11
\tutenum\budapest          % 12
\tutenum\mit               % 13
\tutenum\panjab            % 14 
\tutenum\debrecen          % 15
\tutenum\dublin            % 16
\tutenum\florence          % 17
\tutenum\cern              % 18
\tutenum\wl                % 19
\tutenum\geneva            % 20
\tutenum\hefei             % 21
\tutenum\lausanne          % 22
\tutenum\lyon              % 23
\tutenum\madrid            % 24
\tutenum\florida           % 25
\tutenum\milan             % 26
\tutenum\moscow            % 27
\tutenum\naples            % 29
\tutenum\cyprus            % 30
\tutenum\nymegen           % 31
\tutenum\caltech           % 32
\tutenum\perugia           % 33
\tutenum\peters            % 34
\tutenum\cmu               % 35
\tutenum\potenza           % 36
\tutenum\prince            % 37
\tutenum\riverside         % 38
\tutenum\rome              % 39
\tutenum\salerno           % 40
\tutenum\ucsd              % 41
\tutenum\sofia             % 42
\tutenum\korea             % 43
\tutenum\purdue            % 44
\tutenum\psinst            % 45
\tutenum\zeuthen           % 46
\tutenum\eth               % 47
\tutenum\hamburg           % 48
\tutenum\taiwan            % 49
\tutenum\tsinghua          % 50

{
\parskip=0pt
\noindent
{\bf The L3 Collaboration:}
\ifx\selectfont\undefined%  old style font selection
 \baselineskip=10.8pt
 \baselineskip\baselinestretch\baselineskip
 \normalbaselineskip\baselineskip
 \ixpt
\else%                      new style font selection
 \fontsize{9}{10.8pt}\selectfont
\fi
\medskip
\tolerance=10000
\hbadness=5000
\raggedright
\hsize=162truemm\hoffset=0mm
\def\r{\rlap,}
\noindent

P.Achard\r\tute\geneva\ 
O.Adriani\r\tute{\florence}\ 
M.Aguilar-Benitez\r\tute\madrid\ 
J.Alcaraz\r\tute{\madrid}\ 
G.Alemanni\r\tute\lausanne\
J.Allaby\r\tute\cern\
A.Aloisio\r\tute\naples\ 
M.G.Alviggi\r\tute\naples\
H.Anderhub\r\tute\eth\ 
V.P.Andreev\r\tute{\lsu,\peters}\
% ...
\newpage
%\rule{\textwidth}{0.4pt}
\begin{list}{A}{\itemsep=0pt plus 0pt minus 0pt\parsep=0pt plus 0pt minus 0pt
                \topsep=0pt plus 0pt minus 0pt}
\item[\aachen]
 III. Physikalisches Institut, RWTH, D-52056 Aachen, Germany$^{\S}$
\item[\nikhef] National Institute for High Energy Physics, NIKHEF, 
     and University of Amsterdam, NL-1009 DB Amsterdam, The Netherlands
\item[\mich] University of Michigan, Ann Arbor, MI 48109, USA
\item[\lapp] Laboratoire d'Annecy-le-Vieux de Physique des Particules, 
     LAPP,IN2P3-CNRS, BP 110, F-74941 Annecy-le-Vieux CEDEX, France
\item[\basel] Institute of Physics, University of Basel, CH-4056 Basel,
     Switzerland
\item[\lsu] Louisiana State University, Baton Rouge, LA 70803, USA
\item[\beijing] Institute of High Energy Physics, IHEP, 
  100039 Beijing, China$^{\triangle}$ 
\item[\bologna] University of Bologna and INFN-Sezione di Bologna, 
     I-40126 Bologna, Italy
\item[\tata] Tata Institute of Fundamental Research, Mumbai (Bombay) 400 005, India
\item[\ne] Northeastern University, Boston, MA 02115, USA
\item[\bucharest] Institute of Atomic Physics and University of Bucharest,
     R-76900 Bucharest, Romania
\item[\budapest] Central Research Institute for Physics of the 
     Hungarian Academy of Sciences, H-1525 Budapest 114, Hungary$^{\ddag}$
\item[\mit] Massachusetts Institute of Technology, Cambridge, MA 02139, USA
\item[\panjab] Panjab University, Chandigarh 160 014, India.
\item[\debrecen] KLTE-ATOMKI, H-4010 Debrecen, Hungary$^\P$
\item[\dublin] Department of Experimental Physics,
  University College Dublin, Belfield, Dublin 4, Ireland
\item[\florence] INFN Sezione di Firenze and University of Florence, 
     I-50125 Florence, Italy
\item[\cern] European Laboratory for Particle Physics, CERN, 
     CH-1211 Geneva 23, Switzerland
\item[\wl] World Laboratory, FBLJA  Project, CH-1211 Geneva 23, Switzerland
\item[\geneva] University of Geneva, CH-1211 Geneva 4, Switzerland
\item[\hefei] Chinese University of Science and Technology, USTC,
      Hefei, Anhui 230 029, China$^{\triangle}$
\item[\lausanne] University of Lausanne, CH-1015 Lausanne, Switzerland
\item[\lyon] Institut de Physique Nucl\'eaire de Lyon, 
     IN2P3-CNRS,Universit\'e Claude Bernard, 
     F-69622 Villeurbanne, France
\item[\madrid] Centro de Investigaciones Energ{\'e}ticas, 
     Medioambientales y Tecnol\'ogicas, CIEMAT, E-28040 Madrid,
     Spain${\flat}$ 
\item[\florida] Florida Institute of Technology, Melbourne, FL 32901, USA
\item[\milan] INFN-Sezione di Milano, I-20133 Milan, Italy
\item[\moscow] Institute of Theoretical and Experimental Physics, ITEP, 
     Moscow, Russia
\item[\naples] INFN-Sezione di Napoli and University of Naples, 
     I-80125 Naples, Italy
\item[\cyprus] Department of Physics, University of Cyprus,
     Nicosia, Cyprus
\item[\nymegen] University of Nijmegen and NIKHEF, 
     NL-6525 ED Nijmegen, The Netherlands
\item[\caltech] California Institute of Technology, Pasadena, CA 91125, USA
\item[\perugia] INFN-Sezione di Perugia and Universit\`a Degli 
     Studi di Perugia, I-06100 Perugia, Italy   
\item[\peters] Nuclear Physics Institute, St. Petersburg, Russia
\item[\cmu] Carnegie Mellon University, Pittsburgh, PA 15213, USA
\item[\potenza] INFN-Sezione di Napoli and University of Potenza, 
     I-85100 Potenza, Italy
\item[\prince] Princeton University, Princeton, NJ 08544, USA
\item[\riverside] University of Californa, Riverside, CA 92521, USA
\item[\rome] INFN-Sezione di Roma and University of Rome, ``La Sapienza",
     I-00185 Rome, Italy
\item[\salerno] University and INFN, Salerno, I-84100 Salerno, Italy
\item[\ucsd] University of California, San Diego, CA 92093, USA
\item[\sofia] Bulgarian Academy of Sciences, Central Lab.~of 
     Mechatronics and Instrumentation, BU-1113 Sofia, Bulgaria
\item[\korea]  The Center for High Energy Physics, 
     Kyungpook National University, 702-701 Taegu, Republic of Korea
\item[\purdue] Purdue University, West Lafayette, IN 47907, USA
\item[\psinst] Paul Scherrer Institut, PSI, CH-5232 Villigen, Switzerland
\item[\zeuthen] DESY, D-15738 Zeuthen, Germany
\item[\eth] Eidgen\"ossische Technische Hochschule, ETH Z\"urich,
     CH-8093 Z\"urich, Switzerland
\item[\hamburg] University of Hamburg, D-22761 Hamburg, Germany
\item[\taiwan] National Central University, Chung-Li, Taiwan, China
\item[\tsinghua] Department of Physics, National Tsing Hua University,
      Taiwan, China
\item[\S]  Supported by the German Bundesministerium 
        f\"ur Bildung, Wissenschaft, Forschung und Technologie
\item[\ddag] Supported by the Hungarian OTKA fund under contract
numbers T019181, F023259 and T037350.
\item[\P] Also supported by the Hungarian OTKA fund under contract
  number T026178.
\item[$\flat$] Supported also by the Comisi\'on Interministerial de Ciencia y 
        Tecnolog{\'\i}a.
\item[$\sharp$] Also supported by CONICET and Universidad Nacional de La Plata,
        CC 67, 1900 La Plata, Argentina.
\item[$\triangle$] Supported by the National Natural Science
  Foundation of China.
\end{list}
}
\vfill

%%% Local Variables: 
%%% mode: latex
%%% TeX-master: t
%%% End:

%
%%%%%%%%%%%%%%%%%%%%%%%%%%%%%%%%%%%%%%%%%%%%%%%%%%%%%%%%%%%%%%%%%%%%%%%%%%%%%
% Tables
%%%%%%%%%%%%%%%%%%%%%%%%%%%%%%%%%%%%%%%%%%%%%%%%%%%%%%%%%%%%%%%%%%%%%%%%%%%%%
%

\begin{table}
\begin{center}
\begin{tabular}{|c|c|c|}                                       \hline
$\sqrt{s}$ (\GeV{}) & Named as & ${\cal L}$ (pb$^{-1}$)  \\ \hline
188.6       & 189 &           176.0  \\
191.6       & 192 & \phantom{1}29.5  \\
195.5       & 196 & \phantom{1}83.9  \\
199.5       & 200 & \phantom{1}81.3  \\
201.7       & 202 & \phantom{1}34.8   \\
202.5$-$205.5 & 205 & \phantom{1}74.8  \\
205.5$-$207.2 & 207 &           130.2  \\ 
207.2$-$209.2 & 208 & \phantom{11}8.6  \\ \hline
\end{tabular}
\end{center}

\icaption{Centre-of-mass energies, naming convention and corresponding
 integrated luminosities.
\label{tab:lumeff}}
\end{table} 

\begin{table}
  \begin{center}
    \begin{tabular}{|c|c|c|c|c|c|c|} \hline
    & \multicolumn{2}{c|}{Single-Photon} 
    & \multicolumn{2}{c|}{Single-Photon} 
    &  \multicolumn{2}{c|}{Multi-Photon } \\ 
    & \multicolumn{2}{c|}{$P_t^\gamma > 0.02 \sqrt{s}$ } 
    & \multicolumn{2}{c|}{$P_t^\gamma < 0.02 \sqrt{s}$  }
    & \multicolumn{2}{c|}{$P^{\gamma\gamma}_t > 0.02 \sqrt{s}$}\\
    & \multicolumn{2}{c|}{ } 
    & \multicolumn{2}{c|}{$P_t^\gamma> 0.008 \sqrt{s}$  }
    & \multicolumn{2}{c|}{$E_{\gam}>1 \gev$}\\
    \cline{2-7}
    $\sqrt{s}$ (\GeV{})
    & \multicolumn{1}{c|}{Data} 
    & \multicolumn{1}{c|}{Expected} 
    & \multicolumn{1}{c|}{Data} 
    & \multicolumn{1}{c|}{Expected} 
    & \multicolumn{1}{c|}{Data} 
    & \multicolumn{1}{c|}{Expected} \\ 
    \hline
      189 & \phantom{0}607 & \phantom{0}615.6 &           160 &           162.2 & \phantom{0}26 & \phantom{0}36.2 \\ 
      192 & \phantom{00}89 & \phantom{00}94.6 & \phantom{0}34 & \phantom{0}29.9 & \phantom{0}11 & \phantom{00}5.8 \\ 
      196 & \phantom{0}256 & \phantom{0}258.4 & \phantom{0}79 & \phantom{0}84.7 & \phantom{0}17 & \phantom{0}15.6 \\ 
      200 & \phantom{0}241 & \phantom{0}238.3 & \phantom{0}77 & \phantom{0}80.3 & \phantom{0}15 & \phantom{0}15.0 \\ 
      202 & \phantom{0}114 & \phantom{0}102.0 & \phantom{0}35 & \phantom{0}36.4 & \phantom{00}3 & \phantom{00}6.2 \\ 
      205 & \phantom{0}213 & \phantom{0}210.1 & \phantom{0}74 & \phantom{0}64.7 & \phantom{0}10 & \phantom{0}12.6 \\ 
      207 & \phantom{0}354 & \phantom{0}362.5 & \phantom{0}98 &           112.2 & \phantom{0}17 & \phantom{0}22.0 \\ 
      208 & \phantom{00}24 & \phantom{00}23.5 & \phantom{00}9 & \phantom{00}7.4 & \phantom{00}2 & \phantom{00}1.5 \\ \hline
      Total & 1898 & 1905.1 & 566 & 577.8 & 101 & 114.8  \\ \hline  
    \end{tabular}
   \end{center}
  \icaption{Numbers of observed and expected events selected in
different kinematic regions for different values of $\rts$.
\label{tab:sgamsel}}
\end{table}

\begin{table}
\begin{center}
\begin{tabular}{|c|cccccc|}
\cline{2-7}
\multicolumn{1}{c}{} & \multicolumn{6}{|c|}{$M_{rec}$ [\GeV]} \\ 
\cline{1-1}
{ $| \cos\theta_\gamma |$} &   $0-70$   &   $70-95$   &   $95-120$  &   $120-145$   &   $145-170$  &  $170-210$ \\ 
\hline
$0.000-0.200$ & 
  1/0.5/82 &
 \phantom{0}55/52.9/88 &
 34/38.5/87 &
 18/16.8/88 &
 26/23.6/82 &
 66/74.8/73 \\ 
$0.200-0.400$  &   
  1/0.5/80 &
  \phantom{0}48/65.5/89 &
 49/40.1/89 &
 31/16.8/85 &
 22/25.6/84 &
 93/79.2/73 \\ 
$0.400-0.600$ & 
  0/0.4/81 &
  \phantom{0}67/81.8/88 &
 57/54.9/88 &
 24/22.2/87 &
 33/32.2/83 &
 91/90.0/73 \\ 
$0.600-0.730$ &
 0/0.6/79 &
  \phantom{0}82/68.2/84 &
 44/54.2/84 &
 27/19.9/83 &
 26/29.2/81 &
76/68.7/68 \\ 
$0.800-0.870$ & 
  0/0.7/80 &
  \phantom{0}82/83.0/93 &
 59/60.2/93 &
 28/26.2/91 &
 24/31.2/85 &
 66/58.7/47 \\ 
$0.870-0.920$ & 
 0/0.7/76 &
100/91.9/91 &
 61/65.9/90 &
 26/25.5/86 &
 30/32.8/78 &
51/50.4/37 \\ 
$0.920-0.953$ & 
  0/0.5/60 &
  \phantom{0}94/97.3/87 &
 61/69.9/84 &
 28/24.7/79 &
 20/24.9/57 &
 31/32.8/22 \\ 
 $0.953-0.972$ & 
 0/0.3/59 &
  \phantom{0}82/78.9/70 &
 47/52.7/68 &
 24/20.4/64 &
 12/16.5/36 &
  \phantom{0}1/\phantom{0}2.2/\phantom{0}3 \\ 
\hline
\end{tabular} 
\end{center}
\icaption{Numbers of events selected by the high energy single-photon
 selection, Standard Model expectations and selection efficiencies in \% as a
 function of the recoil mass, $M_{rec}$, and of the photon polar
 angle, $|\cos\theta_\gamma|$. The phase space region corresponding
to this selection is defined in the text.
 \label{tab:highener_sg}}
\end{table}

\begin{table}
\begin{center}
\begin{tabular}{|c|cccccc|}
\cline{2-7}
\multicolumn{1}{c}{} & \multicolumn{6}{|c|}{$M_{rec}$ [\GeV]} \\ 
\cline{1-1}
 \multicolumn{1}{|c|}{$E_{\gamma 2}$ [\GeV]} &   $0-70$   &   $70-95$   &   $95-120$   &   $120-150$
  &   $150-180$ & $180-210$ \\ 
\hline
\multicolumn{1}{|c}{} & \multicolumn{6}{c|}{Full sample} \\ 
\hline
$\phantom{0}0-15$ & 
 0/0.2/59 &
 34/30.6/60 &
 19/21.1/61 &
  9/10.3/58 &
 13/17.6/54 &
  7/7.4/39  \\ 
$15-40$  &   
 0/0.1/64 &
 12/12.4/52 &
\phantom{0}5/\phantom{0}8.2/55 &
2/\phantom{0}3.2/54 &
\phantom{0}0/\phantom{0}0.9/59 &
 ---  \\ 
$40-80$ &
 0/0.2/62 &
  \phantom{0}0/\phantom{0}1.9/60 &
  \phantom{0}0/\phantom{0}0.5/54 &
  --- &
  --- &
  --- \\
\hline
\multicolumn{1}{|c}{} & \multicolumn{6}{c|}{Both Photons in $43^\circ < \theta_\gamma<137^\circ$} \\ 
\hline
$\phantom{0}0-15$ & 
  0/0.1/74 &
  5/6.0/71 &
  4/4.7/78 &
  2/2.1/69 &
  2/4.5/65 &
  1/2.1/45 \\ 
$15-40$  &   
  0/0.0/75 &
  6/3.2/69 &
  1/2.1/77 &
  0/1.0/80 &
  0/0.3/75 &
  --- \\ 
$40-80$ &
  0/0.2/68 &
  0/0.7/73 &
  0/0.1/75 &
  --- &
  --- &
  --- \\
\hline
\end{tabular} 
\end{center}
\icaption{Numbers of observed and expected 
multi-photon events  and
 selection efficiencies in \% as a function of $M_{rec}$ and $ E_{\gamma 2}$
 for the full sample and for the case in which both photons are in the
 barrel. The phase space region corresponding
to the  multi-photon selection is defined in the text.
 \label{tab:multiphoton}}
\end{table}

\begin{table}
\begin{center}
\rotatebox{90}{
\begin{tabular}{|c|lr|lr|lr|lr|lr|lr|lr|}
\cline{2-15}
\multicolumn{1}{c}{} & \multicolumn{14}{|c|}{$x_\gam$} \\ 
\cline{1-1}
\multicolumn{1}{|c|}{$|\cos\theta_\gamma|$}
& \multicolumn{2}{c}{$0.00-0.02$} 
& \multicolumn{2}{c}{$0.02-0.03$} 
& \multicolumn{2}{c}{$0.03-0.05$}
& \multicolumn{2}{c}{$0.05-0.10$} 
& \multicolumn{2}{c}{$0.10-0.20$} 
& \multicolumn{2}{c}{$0.20-0.35$} 
& \multicolumn{2}{c|}{$0.35-0.50$}   \\ 
\hline
\raisebox{-0.6em}{$0.00-0.20$} & 
 29 & 19.8 &
 39 & 39.5 &
 25 & 20.7 &
 28 & 28.5 &
 22 & 29.7 &
 24 & 22.5 &
 13 & 14.5 \\ 
 & 
 28 & 17 &
 54 & 31 &
 64 & 86 &
 68 & 99 &
 79 & 99 &
 82 & 99 &
 83 & 99 \\
\hline
\raisebox{-0.6em}{$0.20-0.40$}  &   
 31 & 30.3 &
 57 & 52.8 &
 27 & 23.8 &
 36 & 29.4 &
 36 & 32.0 &
 20 & 25.8 &
 17 & 15.1 \\
 & 
 33 & 11 &
 53 & 24 &
 63 & 83 &
 68 & 99 &
 79 & 99 &
 83 & 99 &
 84 & 99 \\ 
\hline
\raisebox{-0.6em}{$0.40-0.60$} & 
 19&  17.3 &
111& 105.9 &
 55&  57.4 &
 36&  36.8 &
 44&  37.6 &
 28&  30.4 &
 21&  19.7\\ 
 & 
 36 & 11 &
 50 & 13 &
 63 & 41 &
 67 & 97 &
 78 & 98 &
 83 & 99 &
 84 & 99 \\ 
\hline
\raisebox{-0.6em}{$0.60-0.73$} &
  \multicolumn{2}{c|}{---}  &
111 &135.8 &
 83 & 90.7 &
 27 & 28.1 &
 34 & 32.3 &
 34 &  27.0 &
 17 & 18.0 \\ 
 & 
   \multicolumn{2}{c|}{---}  &
 51 &  8 &
 59 & 22 &
 57 & 94 &
 73 & 99 &
 79 & 99 &
 81 & 99 \\ 
\hline
\raisebox{-0.6em}{$0.87-0.92$} & 
   \multicolumn{2}{c|}{---}  &
   \multicolumn{2}{c|}{---}  &
   \multicolumn{2}{c|}{---}  &
 12 &  17.8 &
 82 &  67.6 &
 42 &  57.3 &
 50 &  41.9 \\ 
 & 
   \multicolumn{2}{c|}{---}  &
   \multicolumn{2}{c|}{---}  &
   \multicolumn{2}{c|}{---}  &
 17  & 96 &
 73  & 99 &
 78  & 99 &
 84  & 98 \\ 
\hline
\raisebox{-0.6em}{$0.92-0.97$} & 
   \multicolumn{2}{c|}{---}  &
   \multicolumn{2}{c|}{---}  &
   \multicolumn{2}{c|}{---}  &
   \multicolumn{2}{c|}{---}  &
 18 &  23.4 &
 24 &  29.8 &
 31 &  32.9 \\ 
 & 
   \multicolumn{2}{c|}{---}  &
   \multicolumn{2}{c|}{---}  &
   \multicolumn{2}{c|}{---}  &
   \multicolumn{2}{c|}{---}  &
 21 &  94 &
 38 & 100 &
 58 & 100 \\ 
\hline
\end{tabular} 
}
\end{center}
\icaption{Numbers of observed and expected single-photon events, 
together with selection efficiencies and purities in \%
 as a function of  the ratio of the photon energy to the beam energy, $x_\gam$, and $|\cos\theta_\gamma|$. Results from
 the combined high and low energy selections are shown.
The phase space regions corresponding
to these selections are defined in the text. In the first row of each
cell, the left number represents the number of observed events and the
right number the expectations from Standard Model processes. In the
second row of each cell, the left number is the selection efficiency
and the right number the purity.
 \label{tab:low_sg}}

\end{table}

\begin{table}
\begin{center}
\begin{tabular}{|c|c|c|c|}                                                  
\hline 
$\sqrt{s}$ (GeV) & $\varepsilon(\%)$ &  $\sigma_{measured}$ (pb) & 
$\sigma_{expected}$ (pb) \\ \hline
189 & $73.7 \pm 0.2$ & $4.83 \pm 0.19 \pm 0.05 $ & 4.97 \\
192 & $71.0 \pm 0.2$ & $4.75 \pm 0.48 \pm 0.05 $ & 4.77 \\
196 & $70.9 \pm 0.2$ & $4.56 \pm 0.28 \pm 0.05 $ & 4.58 \\
200 & $70.4 \pm 0.2$ & $4.44 \pm 0.28 \pm 0.05 $ & 4.39 \\
202 & $70.4 \pm 0.2$ & $4.73 \pm 0.44 \pm 0.05 $ & 4.37 \\
205 & $70.3 \pm 0.2$ & $4.20 \pm 0.28 \pm 0.05 $ & 4.20 \\
207 & $70.6 \pm 0.2$ & $4.00 \pm 0.21 \pm 0.05 $ & 4.15 \\
208 & $69.8 \pm 0.2$ & $4.29 \pm 0.85 \pm 0.05 $ & 4.12 \\
\hline
\end{tabular}
\end{center}

\icaption{Combined trigger and selection efficiency, $\varepsilon$, and measured,
$\sigma_{measured}$, and expected, $\sigma_{expected}$, cross sections
as a function of $\sqrt{s}$ for the
$\epem\ra\nu\bar{\nu}\gamma(\gamma)$ process in the phase space region
defined in the text.  The statistical uncertainty on the selection
efficiency is quoted.  The first uncertainty on $\sigma_{measured}$ is
statistical, the second systematic.  The theoretical uncertainty on
$\sigma_{expected}$ is 1\%~\cite{KK_theory}.
\label{tab:xsecs}}
\end{table}

\begin{table}
\begin{center}
\begin{tabular}{|l|c|}      \hline
Source & Uncertainty (\%) \\ \hline
Trigger efficiency & 0.6 \\
Monte Carlo modelling & 0.6 \\
Selection of converted photons & 0.5 \\
Photon identification & 0.3 \\
Monte Carlo statistics & 0.3 \\
Luminosity & 0.2 \\
Background level& 0.2 \\
Cosmic contamination & 0.1 \\
Calorimeter calibration & 0.1 \\
 \hline
Total & 1.1 \\
\hline
\end{tabular}
\end{center}
\icaption{
Systematic uncertainties on the measurement of the
$\epem\ra\nu\bar{\nu}\gamma(\gamma)$ cross section.
\label{tab:syst}}
\end{table}

\begin{table}
  \begin{center}
    \begin{tabular}{|c|cc|c|c|c|} \hline 
      $n$ & \multicolumn{2}{|c|}{(1/$M_D$)$^{n+2}$} & $M_{D95}$ (\TeV) &  $M_{exp}$ (\TeV) &
$R_{95}$ (cm)   \\ \hline
   2 & $-0.03 \pm 0.10$ &\TeV$^{-4\phantom{0}}$  & 1.50 & 1.49 & $2.1 \times 10^{-2
\phantom{0}}$  \\ 
   3 & $-0.10 \pm 0.28$ &\TeV$^{-5\phantom{0}}$  & 1.14 & 1.12 & $2.9 \times 10^{-7
\phantom{0}}$  \\ 
   4 & $-0.5 \pm 1.0$   &\TeV$^{-6\phantom{0}}$  & 0.91 & 0.89 & $1.1 \times 10^{-9
\phantom{0}}$   \\ 
   5 & $-2.2 \pm 3.9$   &\TeV$^{-7\phantom{0}}$  & 0.76 & 0.75 & $4.2 \times 10^{-1
1}$ \\
   6 & $-11.2 \pm 17.7$ &\TeV$^{-8\phantom{0}}$  & 0.65 & 0.64 & $4.7 \times 10^{-1
2}$  \\  
   7 & $-67 \pm 87$     &\TeV$^{-9\phantom{0}}$  & 0.57 & 0.56 & $1.0 \times 10^{-1
2}$  \\  
   8 & $-400 \pm 460$   &\TeV$^{-10}$ & 0.51 & 0.51 & $3.2 \times 10^{-13}$  \\ \hline
    \end{tabular}
  \end{center}
    \icaption{
 Fitted values of (1/$M_D$)$^{n+2}$, together with the 
observed, $M_{D95}$, and expected, $M_{exp}$, lower  limits
 on the gravity scale  as a function of the
number of extra dimensions, $n$.
 Upper limits on the size of the extra dimensions, $R_{95}$, are also given.
 All limits are at the 95\% confidence level.
\label{tab:extrad}}
\end{table}

%
%%%%%%%%%%%%%%%%%%%%%%%%%%%%%%%%%%%%%%%%%%%%%%%%%%%%%%%%%%%%%%%%%%%%%%%%%%%%%%
% Figures
%%%%%%%%%%%%%%%%%%%%%%%%%%%%%%%%%%%%%%%%%%%%%%%%%%%%%%%%%%%%%%%%%%%%%%%%%%%%%%
%

\clearpage
\newpage  

\end{document}